\documentclass[preprint,eqsecnum,aps,showpacs]{revtex4}
\usepackage{bm}

\begin{document}

\count255=\time\divide\count255 by 60 \xdef\hourmin{\number\count255}
  \multiply\count255 by-60\advance\count255 by\time
 \xdef\hourmin{\hourmin:\ifnum\count255<10 0\fi\the\count255}

\newcommand{\xbf}[1]{\mbox{\boldmath $ #1 $}}

\newcommand{\sixj}[6]{\mbox{$\left\{ \begin{array}{ccc} {#1} & {#2} &
{#3} \\ {#4} & {#5} & {#6} \end{array} \right\}$}}

\newcommand{\threej}[6]{\mbox{$\left( \begin{array}{ccc} {#1} & {#2} &
{#3} \\ {#4} & {#5} & {#6} \end{array} \right)$}}

\title{Complete Analysis of Baryon Magnetic Moments in $1/N_c$}

\author{Richard F. Lebed}
\email{Richard.Lebed@asu.edu}

\author{Daniel R. Martin}
\email{daniel.martin@asu.edu}


\affiliation{Department of Physics and Astronomy, Arizona State University, 
Tempe, AZ 85287-1504}

\date{April, 2004}


\begin{abstract}
We generate a complete basis of magnetic moment operators for the $N_c
\! = 3$ ground-state baryons in the $1/N_c$ expansion, and compute
and tabulate all associated matrix elements.  We then compare to
previous results derived in the literature and predict additional
relations among baryon magnetic moments holding to subleading order in
$1/N_c$ and flavor SU(3) breaking.  Finally, we predict all unknown
diagonal and transition magnetic moments to $\leq 0.15 \,
\mu_N$ accuracy, and suggest possible experimental measurements to
improve the analysis even further.
\end{abstract}


\maketitle
\thispagestyle{empty}

\newpage
\setcounter{page}{1}

\section{Introduction} \label{sec:Intro}

The generalization of quantum chromodynamics from 3 to $N_c \! > \! 3$
color charges, called large $N_c$ QCD, has opened a path to
substantial progress in understanding strong interactions at both the
formal and phenomenological levels.  Formal successes spring from the
fact that large $N_c$ QCD exhibits a well-defined limit, meaning that
the renormalization group equations remain finite and nontrivial as
$N_c \! \to \! \infty$.  Furthermore, the counting of explicit $N_c$
factors organizes QCD Feynman diagrams into topological classes of
decreasing significance with increasing powers of $1/N_c$, which
defines the $1/N_c$ expansion.  Phenomenological successes build on
these formal $1/N_c$ power-counting results, but add one crucial extra
ingredient: Observables calculated to appear at $O(1/N_c)$ or
$O(1/N_c^2)$ are empirically found to be a factor 3 or 9,
respectively, smaller than corresponding quantities calculated to be
$O(N_c^0)$; this means that even QCD with $N_c$ as small as 3 belongs
to the class of theories for which the $1/N_c$ expansion is
meaningful.  We note only that the literature to date that provides
evidence substantiating these statements has become so extensive, that
nothing short of a review article~\cite{reviews} can do it justice.

Nevertheless, a multitude of problems utilizing the $1/N_c$ expansion,
even for well-known observables, remain unsolved.  In this paper we
focus on one very specific such set, the magnetic moments of the
$u,d,s$ baryons in the ground-state multiplet.  In the case of large
$N_c$, this multiplet consists of a tower of states~\cite{DM}
completely symmetric under combined spin and flavor transformations,
thus providing justification for the group-theoretical aspects of the
old three-flavor SU(6) classification for baryons.  The nonstrange
members of the multiplets in this tower carry the $(I,J)$ quantum
numbers ($\frac 1 2$, $\frac 1 2$), ($\frac 3 2$, $\frac 3 2$),
$\ldots$, ($\frac{N_c}{2}, \frac{N_c}{2}$).  The first $(I,J)$
multiplet represents nucleons, which reside in an SU(3) multiplet that
is an octet for $N_c \!  = 3$; the second represents $\Delta$
resonances in an SU(3) multiplet that is a decuplet for $N_c \! = 3$.
Here we continue to use the SU(3) labels {\bf 8} and {\bf 10}, despite
the fact that the corresponding SU(3) representations are much larger
for $N_c \! > \! 3$~\cite{DJM1}.  Of course, for $N_c \! = 3$ this
tower truncates after the $\Delta$'s.  While the mass of each baryon
is $O(N_c^1)$, mass splittings between two low-lying states in the
tower [{\it i.e.}, $I \! = \!  J \! = \!  O(N_c^0)$] is
$O(1/N_c)$~\cite{Jenk}, supporting the notion of a true degenerate
spin-flavor multiplet.  In fact, it is only because our universe is
somewhat closer to the chiral limit than the large $N_c$ limit that
$\Delta$ and its partners in the SU(3) decuplet are unstable under
strong decays: $m_\pi \! = O(m_{u,d}) \! = O(N_c^0) \! < m_\Delta \! -
\! m_N \! = O(1/N_c)$.

In a complete analysis organized according to $1/N_c$, the whole set
of states in the spin-$\frac 1 2$ {\bf 8} and spin-$\frac 3 2$ {\bf
10} [and SU(3) multiplets associated with spin $\frac 5 2$, $\frac 7
2$, $\ldots$, $\frac{N_c}{2}$, which decouple for $N_c \! = 3$], must
be considered together as a single, completely symmetric spin-flavor
multiplet with $N_c$ fundamental representation (quark) indices; we
continue to denote this multiplet by the old SU(6) label {\bf 56},
although again for $N_c \!  > \! 3$ the dimension of this
representation is much greater.  The instability of spin-3/2 baryons
is taken into account simply by maintaining finite values for $m_\pi$
and $1/N_c$ in the full Hamiltonian.

We hasten to add that magnetic moments for baryons in the {\bf 56}
have been considered in the $1/N_c$ expansion in the past---in fact,
in papers dating back a decade or more.  There are three papers in
particular that have examined these magnetic moments in the $1/N_c$
expansion: Jenkins and Manohar (JM)~\cite{JM}, Luty, March-Russell,
and White (LMRW)~\cite{LMRW}, and Dai, Dashen, Jenkins, and Manohar
(DDJM)~\cite{DDJM}.  Each of these papers contains a scheme for
including a particular set of operators that contribute to magnetic
moments, and each is discussed in detail below, once we establish a
notation to describe the formalism.

In short, however, the essential improvement of the current work over
previous papers is completeness.  Once all relevant baryon states are
assumed to fill a complete spin-flavor multiplet---in this case, the
{\bf 56}---then only a finite number of operators exist with distinct
spin-flavor transformation properties that can generate nonvanishing
baryon bilinears in a Hamiltonian.  This number precisely equals the
number of distinct observables associated with the given quantum
numbers.  For example, the {\bf 56} allows precisely 19 linearly
independent mass operators (those with $\Delta J \! = \! 0$, $\Delta
J^3 \! = \!  0$, $\Delta Y \! = \!  0$, $\Delta I^3 \! = \! 0$, T
even) with distinct spin-flavor properties, corresponding to the
masses of the eight {\bf 8} baryons, the ten {\bf 10} baryons, and the
spin-singlet $\Sigma^0 \Lambda$ mixing.  In the magnetic moment case
($\Delta J \! = \! 1$, $\Delta J^3 \! = \! 0$, $\Delta Y \! = \!  0$,
$\Delta I^3 \! = \!  0$, T odd), one finds 27 linearly independent
operators, corresponding to the eight {\bf 8} baryons, ten {\bf 10}
baryons, the $\Delta J \! = \!  1$ $\Sigma^0 \Lambda$ mixing, and
eight SU(3)-breaking $\Delta J \!  = \! 1$ mixings between states of
the same $I^3$ and $Y$ in the {\bf 8} and {\bf 10}, such as $\Delta^+
p$.  Quite simply, these descriptions represent two complete bases of
a vector space corresponding to a particular class of observable: One
basis is organized in such a way as to give one basis vector for each
observable for a given state, and the other basis is organized
according to quantum numbers of the spin-flavor symmetry.  In such an
analysis, an arbitrary amount of symmetry breaking can be
accommodated.

This approach was used to classify all static observables of the
literal SU(6) {\bf 56} ({\it i.e.}, using only $N_c \! = 3$) in
Ref.~\cite{RLSU6}, with a deeper study of quadrupole moments in
Ref.~\cite{BH}.  It has been used in the $1/N_c$ expansion several
times: for the masses of the {\bf 56}~\cite{JL}, for charge radii and
quadrupole moments~\cite{BL1,BHL,BL2}, for the masses and couplings of
the orbitally-excited baryon multiplet {\bf
70}~\cite{Goity,CCGL,CC,GSS,PS}, and even for the $\Delta \to N
\gamma$ couplings closely related to the $\Delta N$ transition
magnetic moments~\cite{JJM}.

This paper is organized as follows: In Sec.~\ref{sec:Basis} we explain
why the operator expansion $1/N_c$ truncates at a finite order, and
how the complete set of operators may be enumerated.  We compute and
tabulate all the matrix elements of all these operators in
Sec.~\ref{sec:Matel}.  In Sec.~\ref{sec:Results} we compare our
approach to previous ones in the literature [with and without
perturbative flavor SU(3) breaking], derive new relations, fit to all
existing data, and use the results of this fit to predict all
unmeasured moments.  The casual reader uninterested in calculational
details is encouraged to skip directly to Sec.~\ref{sec:Results}.  We
summarize and conclude in Sec.~\ref{sec:Concl}.

\section{Operator Basis} \label{sec:Basis}

Each baryon state belongs to a representation composed of $N_c$ color
fundamental representations combined into a color singlet.  While it
is suggestive to think of each such fundamental representation being
associated with a single current quark, such an identification is not
necessary; in general, each fundamental representation merely
represents an interpolating field whose quantum numbers match those of
a single quark in color, spin, and flavor---each of these in the
fundamental representation of the corresponding group---and which
together exhaust the whole baryon wave function~\cite{BL1}.  In
general, such a field consists of not only a current quark, but gluons
and sea quark-antiquark pair Fock components as well, and indeed may
be thought of as a rigorously defined constituent quark.  We continue
to denote such an interpolating field by the simple label ``quark.''

An arbitrary baryon bilinear, as appearing in the Hamiltonian for
masses, magnetic moments, {\it etc.}, thus carries the quantum numbers
of $N_c$ quarks and $N_c$ antiquarks.  Since {\it fundamental\/}
$\otimes$ {\it antifundamental\/} = {\it adjoint\/} $\oplus$ {\it
singlet} for all SU groups (in this case, each of spin, flavor, and
spin-flavor), each operator that can connect the baryon states may be
decomposed into pieces transforming as products of $0,1,2,\ldots,$ up
to $N_c$ adjoint representations.  In terms of spin-flavor SU($2N_F$),
where $N_F$ is the number of light flavors, the operators comprising
the adjoint are defined:
\begin{eqnarray}
J^i & = & \sum_\alpha q^\dagger_\alpha \left( \frac{\sigma^i}{2}
\otimes \openone \right) q_\alpha , \nonumber \\
T^a & = & \sum_\alpha q^\dagger_\alpha \left( \openone \otimes
\frac{\lambda^a}{2} \right) q_\alpha , \nonumber \\
G^{ia} & = & \sum_\alpha q^\dagger_\alpha \left( \frac{\sigma^i}{2}
\otimes \frac{\lambda^a}{2} \right) q_\alpha ,
\end{eqnarray}
where the index $\alpha$ sums over the $N_c$ quarks, $\sigma^i$ are
the Pauli spin matrices, and $\lambda^a$ are the Gell-Mann flavor
matrices.  Thus, each distinct operator may be written as a monomial
in $J$, $T$, and $G$ of total order $n$, with $0 \! \le n \le N_c$.
Such an operator is termed an {\it n-body operator}.

A large subset of operators constructed in this way are redundant or
give vanishing matrix elements due to group-theoretical constraints.
For example, commutators such as $[J^i, J^j ] \! = \! i \epsilon^{ijk}
J^k$ behave exactly as they do for the underlying $\sigma$ and
$\lambda$ matrices.  Furthermore, some combinations of $J$, $T$, and
$G$ act only on non-symmetric combinations of quarks and hence
annihilate the ground-state wave functions, while yet other
combinations are spin-flavor Casimirs and hence give the same value
for every state of the representation, making them indistinguishable
from the identity operator.  The {\it operator reduction rule\/} for
removing all such extra operators was derived for the {\bf 56} in
Ref.~\cite{DJM2}, and extended to the {\bf 70} in Refs.~\cite{CCGL}.
For the present case with $N_F \! \le \! 3$, the rule states: All
operator products in which two flavor indices are contracted using
$\delta^{ab}$, $d^{\, abc}$, or $f^{abc}$~\footnote{Some operators
containing structure constants, such as $d^{\, acg} d^{\, bch} \{ T^g,
G^{ih} \}$, survive the operator reduction rule~\cite{DJM2}.  However,
when acting upon the completely symmetric {\bf 56}, they split into
pieces that either are redundant with simpler operators, have the
wrong sign under time reversal, or annihilate the {\bf 56}.}, or two
spin indices on $G$'s are contracted using $\delta^{ij}$ or
$\epsilon^{ijk}$, can be eliminated.

None of the preceding reasoning depends specifically upon the $1/N_c$
expansion.  Such $1/N_c$ factors arise from two sources: First, an
$n$-body operator appears in an irreducible diagram in which $n$
quarks are connected by gluons, requiring a minimum of $n\!-\!1$
gluons exchanged; the 't~Hooft scaling $\alpha_s \propto 1/N_c$ then
implies an explicit suppression factor $1/N_c^{n-1}$.  Second, the
combinatorics of quarks inside the baryon permits the matrix element
of $J$, $T$, or $G$ to be as large as $O(N_c^1)$ whenever the
contributions from the $N_c$ quarks add coherently.  However, if the
baryons chosen nevertheless have spins, isospins, and strangeness of
$O(N_c^0)$---as we choose for the spin-$\frac 1 2$ {\bf 8} and
spin-$\frac 3 2$ {\bf 10}---then the matrix elements of $J^{1,2,3}$,
$T^{1,2,3}$, and $N_s \! \equiv \!  \frac 1 3 ( \openone \! - \!
2\sqrt{3} \, T^8 )$ are also $O(N_c^0)$.

The replacement of $T^8$ by $N_s$ in constructing the operator basis
presents a trivial example of what, in Ref.~\cite{CCGL}, is called
``operator demotion.''  Whereas operator reduction rules identify
linear combinations of operators that give precisely zero to all
orders in $1/N_c$ when acting upon all states in a baryon spin-flavor
representation, operator demotion identifies linear combinations of
operators whose matrix elements are a higher order in powers of
$1/N_c$ than those of the component operators, at least for the
observed baryon states.  Such a result can occur since the $J$, $T$,
and $G$ matrix elements in general contain both leading and subleading
contributions in $N_c$.

In summary, a complete accounting of the $1/N_c$ expansion thus
requires one to take into account the ingredients: $i)$ a complete set
of operators under spin-flavor; $ii)$ operator reduction rules to
remove linearly dependent operators; $iii)$ a counting of $N_c$
factors arising both from explicit powers of $\alpha_s$ on one hand
and coherent contributions due to quark combinatorics on the other;
and $iv)$ operator demotions to identify operators whose matrix
elements are linearly dependent at $O(1/N_c^n)$ for some $n$ but are
independent at $O(1/N_c^{n+1})$.

A full analysis of all states in the baryon multiplets for $N_c$ large
and finite, as discussed above, requires the inclusion of up to
$N_c$-body operators. A parallel analysis carried out for $N_c \! = 3$
states, by the same reasoning, requires only up to 3-body operators.
Once the physical $N_c \! = 3$ baryons are identified as states
embedded within the $N_c \! > \! 3$ multiplets, one sees that 4-, 5-,
$\ldots$, $N_c$-body operators do indeed act upon the physical
baryons, but give results linearly dependent on those of lower-order
operators, and therefore may be discarded.

In the case of magnetic moments for the {\bf 56}, we have seen that
there are 27 independent parameters with $\Delta J \! = \! 1$, $\Delta
J^3 \! = \! 0$, $\Delta Y \! = \!  0$, $\Delta I^3 \! = \!  0$, T odd
for $N_c \! = \! 3$.  The conditions $\Delta J \! = \! 1$ and $\Delta
J^3 \! = \! 0$ require that each operator has a single unsummed spin
index $i$, which for definiteness we take to be $i \!  = \! 3$.  T
odd, of course, is the behavior of an angular momentum under time
reversal; as it turns out, this is accomplished automatically because
all operators containing the structure constants $\epsilon^{ijk}$,
$f^{abc}$, or $d^{abc}$ can be eliminated.  The conditions $\Delta Y
\! = \!  0$, $\Delta I^3 \!  = \!  0$ require each unsummed flavor
index $a$ to equal 3 or 8.  The complete set of 27 such operators,
including the demotion $T^8 \to N_s$ appears in Table~\ref{oplist}.
For those cases in which different orderings of component operators
would give different values for matrix elements (such as $J^2$ and
$G^{33}$), the operators are written in a symmetric form.

In fact, a direct calculation shows that no other operator demotions
occur.  Consider the first 6 operators in Table~\ref{oplist}, the
complete set up to and including $O(N_c^0)$.  If the leading
coefficient of each of these operators---at $O(N_c^1)$ for $G^{33}$
and at $O(N_c^0)$ for the other 5---for each of the 27 observables is
collected into a $6 \! \times \! 27$ matrix, then one finds that this
matrix has rank 6: No linear combination of the operators has matrix
elements that are only $O(N_c^{-1})$.  Similarly, no combination of
the 17 operators up to $O(N_c^{-1})$ is demoted to $O(N_c^{-2})$.

\section{Computing Matrix Elements} \label{sec:Matel}

We compute the matrix elements of the 27 operators listed in
Table~\ref{oplist} using only the Wigner-Eckart theorem (or its
variants) and the total spin-flavor symmetry of the {\bf 56} baryon
states.  While the task of computing matrix elements of $n$-body
operators for states containing an arbitrarily large number ($N_c$) of
constituents may naively seem to require a large amount of
group-theoretical technology [{\it e.g.}, SU(6) $9j$ symbols], it
turns out that all of the necessary matrix elements can be reduced to
simple SU(2) spin and isospin Clebsch-Gordan (CG) coefficients, and
nothing worse than an SU(2) $6j$ symbol needs to be computed.  All of
the necessary tools have been developed in Refs.~\cite{BHL,BL2}, but
we present them here for completeness.

We begin by constructing baryon states in the {\bf 56}.  Since the
wave function is completely symmetric under exchange of spin and flavor
quantum numbers of any two quarks, it follows that the collection of
all $N_q$ quarks of any fixed flavor $q$ must be completely symmetric
under spin exchange.  The spin $J_q$ carried by them must therefore
have its ``stretched'' value, $N_q/2$.  

Next, the $u$ quarks and $d$ quarks combine to give a state with $I^3
\! = \! \frac 1 2 (N_u \! - \! N_d) \! = \! J_u \! - \! J_d$.  The
total isospin $I$ is determined by noting that the exchange symmetry
property of the state under $u$-$d$ flavor exchange must precisely
match that of these quarks' spins, in order for the total wave
function to be completely symmetric under spin-flavor.  It follows
that $J_{ud} = I$, where $\mathbf{J}_{ud} \! \equiv {\mathbf J}_u
\! + {\mathbf J}_d$.

In the final step, one simply combines the state of $ud$ total spin
$J_{ud} \! = I$ and isospin quantum numbers $I, I^3$ with the
symmetrized strange quarks carrying total spin $J_s$ to obtain the
complete state with spin eigenvalues $J, J^3$, where ${\bf J} \equiv
{\bf J}_{ud} \!  + {\bf J}_s$:
\begin{eqnarray}
\left| J J^3; \, I I^3 \; (J_u \, J_d \, J_s) \right> & = &
\sum_{J_{ud}^3 , \, J_s^3} \left( \begin{array}{cc}
I & J_s \\ J_{ud}^3 & J_s^3 \end{array} \right| \left.
\begin{array}{c} J \\ J^3 \end{array} \right)
\sum_{J_u^3 , \, J_d^3} \left( \begin{array}{cc}
J_u & J_d \\ J_u^3 & J_d^3 \end{array} \right| \left.
\begin{array}{c} I \\ J_{ud}^3 \end{array} \right) \left| J_u \, J_u^3
\right> \left| J_d \, J_s^3 \right> \left| J_s \, J_s^3 \right> ,
\nonumber \\ & & \label{ket}
\end{eqnarray}
where the parentheses denote CG coefficients.  Now, in order to
compute the matrix elements of any particular operator, one need only
sandwich it between a bra and ket of the form of Eq.~(\ref{ket}) and
use the Wigner-Eckart theorem.

The basic operators $T^3$, $N_s$, $J^3$, and {\bf J}$^2$, acting
diagonally on baryon states, are easy to handle even if they are parts
of more complicated operators.  On the other hand,
\begin{eqnarray}
G^{i8} & = & \frac{1}{2\sqrt{3}} (J^i - 3J_s^i) , \nonumber \\
G^{i3} & = & \frac{1}{2} (J_u^i - J_d^i) ,
\end{eqnarray}
are in general not diagonal and must be handled more carefully.
According to Table~\ref{oplist}, they appear in the forms $G^{38}$ and
$G^{33}$, and as
\begin{eqnarray}
J^i G^{i8} & = & \frac{1}{2\sqrt{3}}
(\mathbf{J}^2 - 3 \, \mathbf{J} \! \cdot \! \mathbf{J}_s) ,
\nonumber \\ J^i G^{i3} & = & \frac{1}{2} \, \mathbf{J} \! \cdot \!
(\mathbf{J}_u \! - \mathbf{J}_d) ,
\end{eqnarray}
which may be simplified by noting that
\begin{eqnarray}
\mathbf{J} \! \cdot \mathbf{J}_s & = & -\frac 1 2
[ (\mathbf{J}-\mathbf{J}_s)^2 - \mathbf{J}^2 \! - \mathbf{J}_s^2 ] =
\frac 1 2 ( \mathbf{J}^2 + \mathbf{J}_s^2 - \mathbf{I}^2 ) ,
\nonumber \\
\mathbf{J} \! \cdot \! (\mathbf{J}_u \! - \mathbf{J}_d) & = &
(\mathbf{J}_u \! + \mathbf{J}_d + \mathbf{J}_s) \! \cdot \!
(\mathbf{J}_u \! - \mathbf{J}_d) = \mathbf{J}_u^2 - \mathbf{J}_d^2 +
\mathbf{J}_s \! \cdot \! (\mathbf{J}_u \! - \mathbf{J}_d) .
\end{eqnarray}
It becomes apparent that only a few nontrivial matrix elements need be
computed.  Denoting the matrix element $\langle \mathbf{J}_\alpha
\! \cdot \mathbf{J}_\beta \rangle$ as $\langle \alpha \beta
\rangle^{(0)}$, where $\alpha$ and $\beta$ are any two quark flavors,
the only nontrivial required matrix elements are $\langle J_u^3
\rangle$, $\langle J_d^3 \rangle$, $\langle J_s^3 \rangle$, $\langle
us \rangle^{(0)}$, and $\langle ds \rangle^{(0)}$.

Even more simplification is possible, because Eq.~(\ref{ket}) depends
on the exchange of $u$ and $d$ quarks only through the second CG
coefficient, and the factor obtained through this exchange is just
$(-1)^{J_u + J_d - I}$.  Of course, the eigenvalues $J_\alpha$, which
simply count one-half the number of quarks of flavor $\alpha$ in these
baryons, remain unchanged from initial to final state.  The same is
true for $I^3 \! = J_u \! - \! J_d$, but the total isospin may change
to a value $I^\prime$.  One thus finds for an operator ${\cal O}$ that
\begin{equation} \label{inversion}
\langle I^\prime I^3 \left| {\cal O} (u \leftrightarrow d) \right| I
I^3 \rangle = (-1)^{I^\prime - I} \langle I^\prime - \! I^3 \left|
{\cal O} \right| I - \! I^3 \rangle .
\end{equation}
Thus, the only matrix elements that need be computed are $\langle
J_u^3 \rangle$, $\langle J_s^3 \rangle$, and $\langle us
\rangle^{(0)}$.  These were computed in Ref.~\cite{BL2} and are
reproduced here:
\begin{eqnarray}
\langle J_u^3 \rangle & = & \delta_{J^\prime{}^3 J^3}
\delta_{J_u^\prime J_u^{\vphantom{\prime}}}
\delta_{J_d^\prime J_d^{\vphantom{\prime}}}
\delta_{J_s^\prime J_s^{\vphantom{\prime}}}
(-1)^{J-J^\prime+J^3+J_s+I^\prime-I-J_u-J_d} \nonumber \\ & & \times
\sqrt{J_u (J_u+1) (2J_u+1) (2I^\prime+1) (2I+1) (2J^\prime + 1) (2J+1)}
\nonumber \\ & & \times \sixj{J_d}{J_u}{I}{1}{I^\prime}{J_u}
\sixj{J_s}{I}{J}{1}{J^\prime}{I^\prime}
\threej{1}{J^\prime}{J}{0}{J^3}{-J^3},
\end{eqnarray} 
\begin{eqnarray}
\langle J_s^3 \rangle & = & \delta_{J^\prime{}^3 J_3}
\delta_{I^\prime I} \delta_{J_u^\prime J_u^{\vphantom{\prime}}}
\delta_{J_d^\prime J_d^{\vphantom{\prime}}}
\delta_{J_s^\prime J_s^{\vphantom{\prime}}}
(-1)^{1+J^3+J_s+I} \sqrt{J_s (J_s+1) (2J_s+1) (2J^\prime + 1) (2J+1)}
\nonumber \\ & & \times \sixj{I}{J_s}{J}{1}{J^\prime}{J_s}
\threej{1}{J^\prime}{J}{0}{J^3}{-J^3} ,
\end{eqnarray}
\begin{eqnarray}
\langle u s \rangle^{(0)} & = &
\delta_{J^\prime J} \delta_{J^\prime{}^3 J^3}
\delta_{J_u^\prime J_u^{\vphantom{\prime}}}
\delta_{J_d^\prime J_d^{\vphantom{\prime}}}
\delta_{J_s^\prime J_s^{\vphantom{\prime}}}
(-1)^{1+J+J_s-J_u-J_d} \nonumber \\ & & \times
\sqrt{J_u (J_u+1) (2J_u+1) J_s (J_s + 1) (2J_s +1) (2I^\prime+1) (2I+1)}
\nonumber \\ & & \times
\sixj{J_d}{J_u}{I}{1}{I^\prime}{J_u} \sixj{J}{J_s}{I}{1}{I^\prime}{J_s} .
\end{eqnarray}
Note that, in the interest of exhibiting maximal symmetry, the
remaining CG coefficients have been written as $3j$ symbols.  Despite
the fact that a number of their entries are $O(N_c^1)$, all the $3j$
and $6j$ symbols of interest may be computed using analytic forms
appearing in the standard text by Edmonds~\cite{edmonds}.

The matrix elements for all relevant states are presented in
Tables~\ref{ab0}--\ref{S2f}.  Tables~\ref{ab0}--\ref{S1} are lifted
directly from Ref.~\cite{BL2} (except for the repair of typos in the
$\Sigma^0 \Lambda$ matrix elements in Table~\ref{S1}).

\section{Results} \label{sec:Results}

If data existed for all of the 27 observables associated with the
magnetic moment sector, one would proceed to form a Hamiltonian
\begin{equation}
H = - \bm{\mu} \cdot \mathbf{B} ,
\end{equation}
where the operators ${\cal O}_i$ of Table~\ref{oplist} enter with
unknown dimensionless coefficients $c_i$ via
\begin{equation}
\mu_z = \mu_0 \sum_{i=1}^{27} c_i \, {\cal O}_i ,
\end{equation}
where $\mu_0$ is the sole scale in the problem, a mean value of
magnetic moments in the multiplet, which one expects to be some $O(1)$
multiple of the nuclear magneton $\mu_N$.  The $1/N_c$ expansion
provides a reliable effective Hamiltonian if the coefficients $c_i$
are not larger than $O(1)$.  In fact, a number of them may be smaller
than $O(1)$ because certain operators may only contribute once SU(3)
flavor symmetry is broken.  They may also be smaller if dynamical
effects are present that suppress them below the level predicted by
naive $1/N_c$ counting.  With all 27 observables in hand, one would
simply invert the $27 \! \times \! 27$ matrix whose elements are given
in Tables~\ref{S2a}--\ref{S2f} and solve for all $c_i$ to test this
hypothesis.  Essentially this procedure was carried out for the masses
of the {\bf 56} in Ref.~\cite{JL}.

However, the {\it Review of Particle Physics}~\cite{PDG} gives
unambiguous values for only 10 of the observables: magnetic moments of
7 of the 8 octet baryons ($\mu_{\Sigma^0}$ is unknown), the
$\Omega^-$, and the $\Sigma^0 \Lambda$ and $\Delta^{\!  +} p$
transition moments.  The last of these is extracted from the $\Delta
\! \to \! N \gamma$ helicity amplitudes $A_{\frac 1 2}$ and $A_{\frac
3 2}$ via the standard formula for the M1 amplitude:
\begin{equation}
\mu_{\Delta^{\! +} p} = - \frac{A_{\frac 1 2} + \sqrt{3} A_{\frac 3
2}}{\sqrt{\pi \alpha k}} ,
\end{equation}
where $k \! \simeq \! 260$ MeV is the photon momentum, from which one
finds $\mu_{\Delta^{\! +} p} \! = \! 3.51 \pm 0.09 \, \mu_N$.  In
addition, we use a recent extraction~\cite{LCM} $\mu_{\Delta^{\! ++}}
\! = 6.14 \pm 0.51 \, \mu_N$ obtained from an analysis of data that
has some model dependence, but that respects both gauge invariance and
the finite $\Delta^{\! ++}$ width.  We therefore include 11
observables in our analysis.  There is also a recent experimental
determination~\cite{Kot} of $\mu_{\Delta^{\! +}} \! = 2.7^{+
1.0}_{-1.3} \: {\rm (stat)} \pm 1.5 \: {\rm (syst)} \pm 3 \: {\rm
(theor)} \, \mu_N$, but due to the large theoretical uncertainty we
do not use this value in our analysis.

With only 11 pieces of information to study a system of 27
observables, one must resort to using the known quantities to fit the
coefficients at the lowest orders of the $1/N_c$ expansion, and to use
the coefficients so obtained to predict the remaining observables.
One may then proceed either by $i)$ separating the observables into
isoscalar and isovector, as well as $I \! = \! 2$ and 3 isotensor,
combinations, or $ii)$ one may employ the electromagnetic nature of
magnetic moments to construct only operators with a flavor dependence
in proportion to the quark charges (the ``single-photon
ansatz'').  Since both methods have been employed in the literature,
we discuss them each in turn.

\subsection{Analysis in the Isoscalar, Isovector, Isotensor Basis}

The analysis of Ref.~\cite{JM} (JM) separates operators, and the
corresponding combinations of magnetic moments, into $I \! = 0$ and $I
\! = 1$ forms.  Since the maximal isospin appearing in the {\bf 56}
is $\frac 3 2$ (for the $\Delta$), isotensor combinations with $I
\! = 2$ and $I \! = 3$ are also present:
\begin{eqnarray}
I=2: && (\mu_{\Sigma^+} - 2 \mu_{\Sigma^0} + \mu_{\Sigma^-}) , \ \
(\mu_{\Delta^{++}} \! - \mu_{\Delta^+} - \mu_{\Delta^0} +
\mu_{\Delta^-}) ,
\ \ (\mu_{\Sigma^{*+}} - 2 \mu_{\Sigma^{*0}} + \mu_{\Sigma^{*-}}) ,
\nonumber \\
&& (\mu_{\Delta^{\! +} p} - \mu_{\Delta^0 n}) , \ \
(\mu_{\Sigma^{*+} \Sigma^+} - 2 \mu_{\Sigma^{*0} \Sigma^0} +
\mu_{\Sigma^{*-} \Sigma^-}) , \nonumber \\
I=3:
&& (\mu_{\Delta^{++}} \! - 3 \mu_{\Delta^+} + 3 \mu_{\Delta^0} -
\mu_{\Delta^-}) . \label{I2I3}
\end{eqnarray}

The JM analysis introduces a leading-order operator $X^{ia}$, which is
equivalent to the $O(N_c^0)$ part of $G^{ia} \! /N_c$, and a strange
quark spin operator
\begin{equation}
J_s^i \equiv \frac 1 3 (J^i \! - 2\sqrt{3} G^{i8}).
\end{equation}
The JM operator basis then consists of the 6 operators $N_c X^{i3}$
and $N_s X^{i3}$ ($I \!  = 1$), and $J^i$, $J_s^i$, $N_s J^i/N_c$, and
$N_s J_s^i/N_c$ ($I \! = 0$).  Since no combinations of these
operators have $I \! = 2$ or 3, the combinations in Eqs.~(\ref{I2I3})
exactly vanish, giving relations I1--I6 (JM Table~2).

In comparison with our Table~\ref{oplist}, the choice of JM operators
reflects the inclusion of all (2) with $I \! = 1$ at $O(N_c^1)$ and
$O(N_c^0)$, and all (4) with $I \! = 0$ at $O(N_c^0)$ and
$O(N_c^{-1})$.  Since there are (as one may readily count) 10 $I \! =
0$ and 11 $I \!  = 1$ magnetic moment combinations in the {\bf 56}, it
follows that JM predict 6 isoscalar relations (JM Table~3 S1--S6) that
receive only $O(N_c^{-2})$ corrections, and 9 isovector relations (JM
Table~3 V1--V7, V8$_1$, and V9$_1$) that receive only $O(N_c^{-1})$
corrections.  As expected, we confirm these predictions in our basis.

The JM analysis makes no use of the electromagnetic behavior of
magnetic moments, nor of perturbative SU(3) flavor breaking; its
analysis can be said to hold in the presence of arbitrarily large
SU(3) breaking.  Thus, operators are organized solely by the $1/N_c$
power suppression of their matrix elements.  Since 17 operators occur
up to and including $O(N_c^{-1})$ while only 11 moment parameters have
been measured, it is not yet possible to improve upon the numerical
analysis of JM using their same scheme.  One must therefore impose a
physically natural flavor structure on the magnetic moment operators,
a topic to which we next turn.

\subsection{Analysis Using the Single-Photon Ansatz}

Like all electromagnetic multipole moments, magnetic moments are
defined through a particular coupling of a bilinear to an on-shell
photon.  The lowest-order flavor structure of the coupling to the
photon should therefore be such that each quark couples proportionally
to its electric charge.  In particular, in the limit in which all
other sources of SU(3) breaking are suppressed, only operators with
one unsummed flavor index $a$ may appear, and then only in the linear
combination $(a \! = \! 3) + (a \! = \!  8)/\sqrt{3} \equiv (a \! = \!
Q)$.  Specifically, these are the forms
\begin{eqnarray}
Q = T^Q & \equiv & T^3 + \frac{1}{\sqrt{3}} T^8 , \nonumber \\
G^{iQ} & \equiv & G^{i3} + \frac{1}{\sqrt{3}} \, G^{i8} .
\end{eqnarray}

Implicit in this definition of the quark charge matrix $Q$ is that the
quarks assume their usual $N_c \! = 3$ values $q_u \! = \! +\frac 2
3$, $q_d \! = q_s \! = \! -\frac 1 3$.  In terms of SU(3) flavor
hypercharges, $Y_u \! = \! Y_d \! = \! \frac 1 3$, $Y_s \! = \! -\frac
2 3$.  An alternate choice, $q_u = (N_c \! + 1)/(2N_c)$, $q_d = q_s
= (-N_c \! + 1)/(2N_c)$ ($Y_u \! = Y_d \! = \!  1/N_c$, $Y_s \! =
\!  -1 \! + 1/N_c$), has the convenient property that all hadrons then
have the same electric charges and hypercharges for arbitrary $N_c$ as
they do for $N_c \! = 3$.  Moreover, with this choice the chiral
anomalies of the Standard Model (with $N_c$ colors) automatically
cancel.  However, one is also faced with the mysterious prospect of
electromagnetic charges dependent upon the number $N_c$ of QCD
charges.  More significantly, the quantization condition of the
Wess-Zumino term permits only baryon SU(3) representations containing
states with hypercharge $Y \! = \!  N_c/3$~\cite{CohenWZ}; if such
states have $O(N_c^0)$ strange quarks, then the $N_c$-dependent choice
($Y_u \! = \!  Y_d \! = \! 1/N_c$) is disallowed.  For the remainder
of this paper, we assume the usual $N_c$-independent quark charges.

The only operators occurring in the single-photon ansatz with no other
SU(3) breaking ({\it cf.} Table~\ref{oplist}) are
\begin{equation} \label{LO}
{\cal O}_1 \! = G^{3Q}, \ {\cal O}_2 \! = \frac{1}{N_c} Q J^3, \
\tilde {\cal O}_3 \! = \! \frac{1}{N_c^2} \frac 1 2 \{ J^2, G^{3Q} \} , \
{\cal O}_4 \! = \! \frac{1}{N_c^2} J^i G^{iQ} J^3 .
\end{equation}
For any value of $N_c$ it turns out that the combination ${\cal O}_3
\! \equiv \! (\tilde {\cal O}_3 \! - {\cal O}_4)$ vanishes for all
diagonal moments and survives only for transitions.  Since the only
transition moment measured at present is $\mu_{\Delta^{\! +} p}$,
using ${\cal O}_3$ rather than $\tilde {\cal O}_3$ provides a more
incisive test of the expansion when fitting to current data.

In addition, one may perturbatively break SU(3) symmetry by including
effects due to a finite strange quark mass or spin.  We incorporate
such effects by including a parameter $\varepsilon$ that indicates
each instance of breaking of the SU(3) symmetry.  At first blush, one
may suppose that it is proportional to $m_s$, $\varepsilon \simeq 0.3
\approx 1/N_c$, but as we see below such a rigid identification is not
necessary.  The list of additional operators with matrix elements up
to $O(\varepsilon^1 N_c^0)$ reads
\begin{equation} \label{NLO}
\varepsilon {\cal O}_5 \! = \varepsilon q_s J_s^3 , \
\varepsilon {\cal O}_6 \! = \! \frac{\varepsilon}{N_c} N_s G^{3Q} , \
\varepsilon {\cal O}_7 \! = \! \frac{\varepsilon}{N_c} \, Q J_s^3 .
\end{equation}
These forms are obtained by the simple expedient of inserting sources
of SU(3) breaking along the strangeness direction into the operators
of Eq.~(\ref{LO}), and retaining only those with matrix elements up to
$O(\varepsilon^1 N_c^0)$.  The possible substitutions are $J^i \to
\varepsilon J_s^i$, $Q/N_c \to \varepsilon q_s N_s/N_c$, $G^{iQ} \to
\varepsilon q_s J_s^i$, or multiplication of an operator by
$\varepsilon N_s/N_c$; each of the last three of these replacements
also costs a power of $1/N_c$.  In terms of this basis, the analysis
of Ref.~\cite{LMRW} (LMRW) uses the operators ${\cal O}_1$ [LMRW
Eq.~(17)], ${\cal O}_2$ [LMRW Eq.~(21)], the combination
\begin{equation} \label{nonan}
m_s^{1/2} \left[ - \left( 1 + \frac{3}{N_c} \right)
{\cal O}_5 - {\cal O}_6 + {\cal O}_7 \right] ,
\end{equation}
from chiral loop diagrams [LMRW Eq.~(25)], and the operators $m_s
{\cal O}_5$, $m_s {\cal O}_6$, $m_s {\cal O}_7$ from counterterms to
the loop calculation [LMRW Eq.~(29)], 5 independent operators in all.
Note the characteristic nonanalytic $m_s$ behavior in
Eq.~(\ref{nonan}), which suggests that the appropriate SU(3) expansion
parameter $\varepsilon$ might properly scale as $\simeq \! N_c^{-1/2}$
rather than $\simeq \! N_c^{-1}$; we address this issue below.

At the next order, $O(\varepsilon^1 N_c^{-1})$, one finds the 6
operators
\begin{eqnarray}
&&
\varepsilon {\cal O}_8 = \varepsilon q_s \frac{N_s}{N_c} J^3 , \
\varepsilon {\cal O}_9 = \varepsilon \frac{N_s}{N_c^2} Q J^3 , \
\varepsilon {\cal O}_{10} = \frac{\varepsilon}{N_c^2} \frac 1 2 \{
{\bf J} \! \cdot \! {\bf J}_s , G^{3Q} \}, \
\nonumber \\ &&
\varepsilon {\cal O}_{11} = \frac{\varepsilon}{N_c} J_s^j G^{jQ}
J^3, \
\varepsilon {\cal O}_{12} = \frac{\varepsilon}{N_c^2} \frac 1 2 \{
J^j G^{jQ}, J_s^3 \} . \label{NNLO}
\end{eqnarray}
Beyond this collection, the next operators have matrix elements of
$O(\varepsilon^2 N_c^{-1})$ and $O(\varepsilon^1 N_c^{-2})$, but
significantly there are none of $O(\varepsilon^2 N_c^0)$ or
$O(\varepsilon^0 N_c^{-2})$.  As a consequence, whether one takes the
SU(3)-breaking parameter $\varepsilon \! \simeq \! N_c^{-1/2}$ or
$\simeq \!  N_c^{-1}$, the series expansion truncates consistently
after the inclusion of either the set ${\cal O}_1, \ldots, \varepsilon
{\cal O}_7$ (complete to combined orders $\varepsilon^1 N_c^0$ and
$\varepsilon^0 N_c^{-1}$) or the larger set ${\cal O}_1, \ldots,
\varepsilon {\cal O}_{12}$ [complete to $O(\varepsilon^1 N_c^{-1})$].

In this notation, the operators used in Ref.~\cite{DDJM} (DDJM)
consist of [DDJM Eq.~(2.9)]
\begin{equation}
{\cal O}_1, \ {\cal O}_2, \ {\cal O}_3, \ \varepsilon {\cal O}_5, \
\varepsilon {\cal O}_6, \ \varepsilon {\cal O}_7, \ \varepsilon {\cal
O}_8, 
\end{equation}
and the operator $\varepsilon^2 q_s N_s J_s^3 / N_c$ with a
coefficient fixed relative to those of ${\cal O}_6$ and ${\cal O}_7$,
7 independent operators in all, but a somewhat different set than
${\cal O}_1 , \ldots, \varepsilon {\cal O}_7$.  Note that DDJM does
not assign particular powers of $\varepsilon$; DDJM also recognizes
the presence of $O(m_s^{1/2})$ loop corrections, so that statements
regarding the meaning of $\varepsilon$ still apply.

\subsection{A Global Fit}

As we have seen, assigning a particular numerical value to the
SU(3)-breaking parameter $\varepsilon$ can be problematic, owing to
the existence of contributions nonanalytic in $m_s$.  Fortunately, we
have also seen that regardless of whether one takes $\varepsilon \!
\simeq \! N_c^{-1/2}$ or $N_c^{-1}$, the expansion truncates
consistently after the 7 operators in Eqs.~(\ref{LO}), (\ref{NLO}) or
after the 12 operators including Eqs.~(\ref{NNLO}).  Powers of
$\varepsilon$ may simply be left in the coefficients, in which case
the size of SU(3) breaking for each operator may be judged directly
from a fit to data.

Unfortunately, with only 11 measured magnetic moment parameters, only
a fit to the first 7 operators is possible at present.  We therefore
perform a least-squares fit to the coefficients in the expansion
\begin{equation} \label{newexp}
\mu_z = \mu_0 \sum_{i=1}^7 d_i \, {\cal O}_i .
\end{equation}
We choose the scale $\mu_0$ by reasoning that the best known value
among the moments is $\mu_p$, and that there is only one operator, ${\cal
O}_1$, at leading order ($N_c^1$), whose value for the proton is $(N_c
\! + \! 3)/12 \! = \! 1/2$ for $N_c \! = \! 3$.  Therefore, a natural
choice that makes the sole leading-order coefficient $d_1$ of order
unity is to set $\mu_0 = 2\mu_p$ [In alternate choices one may average
over several measured magnetic moments, but this merely renormalizes
all $d_i$ by the same $O(1)$ multiple].  Since the expansion is
truncated by ignoring effects of $O(\varepsilon^1 N_c^{-1})$, one must
include in addition to the statistical uncertainty of each magnetic
moment a ``theoretical uncertainty'' of order $\mu_0 \,
\varepsilon/N_c$.  In fact, the $\chi^2$/d.o.f.\ obtained from a
theoretical uncertainty of $\mu_p / N_c^2 \simeq 0.3 \, \mu_N$, for
example, is only about 0.13, suggesting that the naive theoretical
uncertainty is a gross overestimate.  We find empirically that
choosing it to have a value about $\mu_p/N_c^3 \simeq 0.1 \, \mu_N$
gives a $\chi^2$/d.o.f.\ = 1.05, meaning that the fit is as good as
one might hope for {\em even if all $O(\varepsilon^1 N_c^{-1})$
effects are suppressed, and uncertainties are effectively only
$O(\varepsilon^2 N_c^{-1})$}.  This result far supersedes that
expected from a naive $1/N_c$ expansion.

The fit values for the coefficients are given in Table~\ref{fit}.  One
immediately notes that no coefficients are larger than $O(1)$; had any
of them been so, one would conclude that the $1/N_c$ expansion fails.
But in fact, $d_1$ and $d_3$ are of $O(1)$, while $d_2$ and $d_4$ are
actually consistent with zero.  The SU(3)-breaking coefficients
$d_{4,5,6}$, do indeed have central values about $1/3$ or less, but
only $d_4$ is statistically different from zero.  Such a pattern of
suppression beyond naive $1/N_c$ counting has in fact been seen
before, in the orbitally excited baryons~\cite{CCGL,CC,GSS}.
Moreover, a hint of this effect is visible in the results of DDJM
Table~VI [although their operator basis, and especially their
treatment of SU(3) breaking, is rather different].  We do not
understand the source of this suppression beyond that expected from
$1/N_c$ counting, and find it to be the most intriguing feature of our
analysis.

One may also use the fit values for $d_i$ to predict all of the
remaining 16 magnetic moments to within the theoretical uncertainty;
the results are presented in Table~\ref{pred}.  Note that the recent
$\mu_{\Delta^+}$ measurement easily agrees with our prediction.  An
important point that may not be obvious from this compilation is that,
with only 7 operators included, there remain 20 relations among the
magnetic moments.  Among these are vanishing of the $I \! = \! 2$ and
3 combinations in Eq.~(\ref{I2I3}): The leading-order operators in
Eq.~(\ref{LO}) contain only $I \! = \! 0$ and 1 pieces, while
inserting SU(3) breaking along the strangeness direction induces only
$I \! = \! 0$ corrections.  The combinations in Eqs.~(\ref{I2I3})
receive contributions only from tiny isospin-breaking effects due to
either $(m_u \!  - \! m_d)$ or loop diagrams containing an additional
photon [$O(\alpha/4\pi)$].  Furthermore, $\mu_{\Delta^0}$ vanishes for
all 7 operators when $N_c \! = 3$, and receives contributions only
from SU(3) breaking not solely in the strangeness direction (since
$\Delta^0$ contains no $s$ quarks), which induces an additional
$m_{u,d}/m_s$ suppression factor.  A similar statement holds for all
nonstrange observables: None of them receive a contribution from any
operators beyond ${\cal O}_1, \ldots , {\cal O}_4$; indeed, the famous
SU(6) relation $\mu_n = -2\mu_p/3$ for $N_c \! = 3$ receives a
contribution only from the anomalously suppressed operator ${\cal
O}_2$.

Predictions of the diagonal moments appearing in LMRW Table~I agree
well with the results of Table~\ref{pred}, but ours have smaller
uncertainties due to the larger number of operators and the treatment
of subleading effects as described above.

Since the next order of the expansion contains 12 operators while 11
observables are well known, it is tempting to suppose that just one
more moment measured---say, an improvement on $\mu_{\Delta^+}$---will
permit such a fit.  However, the $I \! = 3$ and $I \! = 2$ relations
among the $\Delta$'s combined with the result $\mu_{\Delta^0} \! = \!
0$, as satisfied by all operators in our list [and any others breaking
SU(3) solely in the strangeness direction], predict that the $\Delta$
magnetic moments are exactly proportional to electric charge for $N_c
\! = 3$: $\mu_{\Delta^{++}} \! = \! 2 \mu_{\Delta^+} \! = \!  -2
\mu_{\Delta^-}$.  Moreover, there is precisely one relation satisfied
by the first 12 operators among the measured moments:
\begin{equation} \label{reln}
\mu_n -\frac 1 4 (\mu_{\Sigma^+} + \mu_{\Sigma^-}) -\frac 3 2
\mu_{\Lambda} -\sqrt{3} \mu_{\Sigma^0 \Lambda} + \mu_{\Xi^0} =
O(\varepsilon^2 N_c^{-1}) ,
\end{equation}
which has a numerical value of $0.22 \pm 0.14 \, \mu_N$, the
uncertainty being completely dominated by that of $\mu_{\Sigma^0
\Lambda}$.  In fact, Eq.~(\ref{reln}) was originally derived in heavy
baryon chiral perturbation theory~\cite{JLSM}, where it was found to
have no $O(m_s^{1/2})$, $O(m_s)$, or $O(m_s \! \ln m_s)$
corrections---{\it i.e.}, no $O(\varepsilon^1)$ corrections in the
current formalism.  Converting Eq.~(\ref{reln}) into a scale-invariant
result by dividing by the average of the same expression with all
negative values turned to positive ones [giving an $O(N_c^1)$
combination], one obtains $0.057 \pm 0.036$, in good agreement with
expected magnitude $\varepsilon^2 N_c^{-2}$.  A better measurement of
$\mu_{\Sigma^0 \Lambda}$, certainly within current experimental means,
would decisively test the expansion at this order.

It follows that a measurement of at least two moments with nonzero
strangeness is required to perform a fit to the 12 operators at
$O(\varepsilon^1 N_c^{-1})$, and to determine whether the effects at
this order are truly suppressed, as the $\chi^2$/d.o.f.\ suggests.
The $\Sigma^* \Lambda$, $\Sigma^* \Sigma$, and $\Xi^* \Xi$ transitions
are natural candidates, since the associated radiative decays are
presumably being recorded (although not yet studied) at Jefferson Lab,
as well as other facilities.  To date, the decay $\Sigma^{*0} \! \to
\Lambda \gamma$ has been seen in precisely one event~\cite{Meis72}; the
opportunity to improve on this meager set clearly exists.

Finally, we note in passing that once the unmeasured moments are
included, a number of relations with only $O(\varepsilon^2 N_c^{-1})$
corrections, in addition to Eqs.~(\ref{I2I3}), (\ref{reln}), and
$\mu_{\Delta^0} \! = \! 0$, remain (7, to be precise).  A particularly
elegant example is $\mu_{\Sigma^{*-} \Sigma^-} \! = \mu_{\Xi^{*-}
\Xi^-}$.

\section{Conclusions} \label{sec:Concl}

We have developed a basis of operators representing every possible
observable pattern of magnetic moments for the ground-state spin-1/2
and spin-3/2 baryon multiplets, and organized them according to the
counting of $1/N_c$ factors.  We have furthermore computed the
group-theoretical parts of all of these operators, thus producing a
complete effective Hamiltonian for magnetic moments.  Our analysis of
this operator expansion examined the consequences both in the case of
arbitrarily large SU(3) breaking and perturbative SU(3) breaking (in
powers of a parameter $\varepsilon$) beyond that produced by coupling
quarks to photons in proportion to their electric charges.  In both
cases we have compared to previous results and showed how this work
extends earlier analyses.

In particular, we have found in the case of nonperturbative SU(3)
breaking that the measurement of several additional magnetic moments
is necessary to improve numerically upon previous analysis [{\it
i.e.}, from relative order $N_c^{-2}$ to $N_c^{-3}$].  However, in the
more physically meaningful case of perturbative SU(3) breaking, the
series may be truncated consistently after 7 operators (including up
to orders $\varepsilon^1 N_c^0$ and $\varepsilon^0 N_c^{-1}$) or after
12 operators [up to $O(\varepsilon^1 N_c^{-1})$].  Since 11
observables are currently well measured, we presented results of a fit
to 7 operators, and found not only that several of the effective
Hamiltonian coefficients are smaller than expected, but also that a
good fit can be obtained if the terms neglected are actually $1/N_c$
smaller than naively expected.

A number of relations among the magnetic moments survive the expansion
to 12 operators.  After enumerating a number of them [{\it e.g.},
Eq.~(\ref{reln})], we suggested that the most probative tests of the
$1/N_c$ expansion: an improved measurement of the $\Sigma^0 \Lambda$
transition and observation of $\Sigma^* \Lambda$, $\Sigma^* \Sigma$,
and $\Xi^* \Xi$ transitions, should lie with current experimental
means.

\section*{Acknowledgments}
This work was supported in part by the National Science Foundation
under Grant No.\ PHY-0140362.

\clearpage
\begin{table}
\caption{The 27 linearly independent operators contributing to the
magnetic moments of the spin-$\frac 1 2$ and spin-$\frac 3 2$
ground-state baryons, organized according to the leading $N_c$
counting of their matrix elements.\label{oplist}}
\begin{tabular}{r|l}
\hline\hline
$O(N_c^1)$ \ & \ $G^{33}$ \\
$O(N_c^0)$ \ & \ $J^3$, \ $G^{38}$, \ $\frac{1}{N_c} T^3 G^{33}$, \
$\frac{1}{N_c} N_s G^{33}$, \ $\frac{1}{N_c^2} \frac 1 2 \{ J^i G^{i3},
G^{33} \}$ \\
$O(N_c^{-1})$ \ & \ $\frac{1}{N_c} T^3 J^3$, \ $\frac{1}{N_c} N_s J^3$,
\ $\frac{1}{N_c} T^3 G^{38}$, \ $\frac{1}{N_c} N_s G^{38}$, \
$\frac{1}{N_c^2} \frac 1 2 \{ J^2, G^{33} \}$, \ $\frac{1}{N_c^2}
(T^3)^2 G^{33}$, $\frac{1}{N_c^2} N_s^2 G^{33}$, \\
& \ $\frac{1}{N_c^2} T^3 N_s G^{33}$, \ $\frac{1}{N_c^2} J^i G^{i3}
J^3$, \ $\frac{1}{N_c^2} \frac 1 2 \{ J^i G^{i8}, G^{33} \}$, \
$\frac{1}{N_c^2} \frac 1 2 \{ J^i G^{i3}, G^{38} \}$ \\
$O(N_c^{-2})$ \ & \ $\frac{1}{N_c^2} J^2 J^3$, \
$\frac{1}{N_c^2} N_s^2 J^3$, \ $\frac{1}{N_c^2} (T^3)^2 J^3$, \
$\frac{1}{N_c^2} T^3 N_s J^3$, \
$\frac{1}{N_c^2} \frac 1 2 \{ J^2, G^{38} \}$, \
$\frac{1}{N_c^2} (T^3)^2 G^{38}$, $\frac{1}{N_c^2} N_s^2 G^{38}$, \\
& \ $\frac{1}{N_c^2} T^3 N_s G^{38}$, \
$\frac{1}{N_c^2} J^i G^{i8} J^3$, \
$\frac{1}{N_c^2} \frac 1 2 \{ J^i G^{i8}, G^{38} \}$ \\
\hline
\end{tabular}
\end{table}

\begin{table}
\caption{Matrix elements of the operators $N_{u,d,s}$ [whence $\langle
\mathbf{J}_\alpha^2 \rangle = \langle \alpha \alpha \rangle^{(0)} =
(N_\alpha/2)(N_\alpha/2 + 1)$] and the rank-0 tensors $\langle
\alpha \beta \rangle^{(0)}$ with $\alpha \neq \beta$.  Since spin is
unchanged by these operators, the matrix elements vanish for all
off-diagonal transitions except $\Sigma^0 \Lambda$; in that case, the
only nonvanishing entries are $\langle u s \rangle^{(0)} = -\langle d s
\rangle^{(0)} = -\frac{1}{8}\sqrt{(N_c-1)(N_c+3)}$.\label{ab0}}
\begin{tabular}{c|c|c|c|c|c|c}
\hline\hline
State & $\langle N_u \rangle$ & $\langle N_d \rangle$ & $\langle N_s
\rangle$ & $\langle u d \rangle^{(0)}$ & $\langle u s \rangle^{(0)}$ &
$\langle d s \rangle^{(0)}$ \\
\hline\hline
$\Delta^{++}$ & $\frac{1}{2}(N_c+3)$ & $\frac{1}{2}(N_c-3)$ & 0 &
$-\frac{1}{16} (N_c-3)(N_c+7)$ & 0 & 0 \\
$\Delta^+$ & $\frac{1}{2}(N_c+1)$ & $\frac{1}{2}(N_c-1)$ & 0 &
$-\frac{1}{16} (N_c^2 + 4N_c - 29)$ & 0 & 0 \\
$\Delta^0$ & $\frac{1}{2}(N_c-1)$ & $\frac{1}{2}(N_c+1)$ & 0 &
$-\frac{1}{16} (N_c^2 + 4N_c - 29)$ & 0 & 0 \\
$\Delta^-$ & $\frac{1}{2}(N_c-3)$ & $\frac{1}{2}(N_c+3)$ & 0 &
$-\frac{1}{16} (N_c-3)(N_c+7)$ & 0 & 0 \\
$\Sigma^{*+}$ & $\frac{1}{2}(N_c+1)$ & $\frac{1}{2}(N_c-3)$ & 1 &
$-\frac{1}{16} (N_c-3)(N_c+5)$ & $+\frac{1}{16}(N_c+5)$ &
$-\frac{1}{16}(N_c-3)$ \\
$\Sigma^{*0}$ & $\frac{1}{2}(N_c-1)$ & $\frac{1}{2}(N_c-1)$ & 1 &
$-\frac{1}{16} (N_c^2 + 2N_c - 19)$ & $+\frac 1 4 $ & $+\frac 1 4$ \\
$\Sigma^{*-}$ & $\frac{1}{2}(N_c-3)$ & $\frac{1}{2}(N_c+1)$ & 1 &
$-\frac{1}{16} (N_c-3)(N_c+5)$ & $-\frac{1}{16}(N_c-3)$ &
$+\frac{1}{16}(N_c+5)$ \\
$\Xi^{*0}$ & $\frac{1}{2}(N_c-1)$ & $\frac{1}{2}(N_c-3)$ & 2 &
$-\frac{1}{16} (N_c-3)(N_c+3)$ & $+\frac{1}{12}(N_c+3)$ &
$-\frac{1}{12}(N_c-3)$ \\
$\Xi^{*-}$ & $\frac{1}{2}(N_c-3)$ & $\frac{1}{2}(N_c-1)$ & 2 &
$-\frac{1}{16} (N_c-3)(N_c+3)$ & $-\frac{1}{12}(N_c-3)$ &
$+\frac{1}{12}(N_c+3)$ \\
$\Omega^-$ & $\frac{1}{2}(N_c-3)$ & $\frac{1}{2}(N_c-3)$ & 3 &
$-\frac{1}{16} (N_c-3)(N_c+1)$ & 0 & 0 \\
$p$ & $\frac{1}{2}(N_c+1)$ & $\frac{1}{2}(N_c-1)$ & 0 & $-\frac{1}{16}
(N_c-1)(N_c+5)$ & 0 & 0 \\
$n$ & $\frac{1}{2}(N_c-1)$ & $\frac{1}{2}(N_c+1)$ & 0 & $-\frac{1}{16}
(N_c-1)(N_c+5)$ & 0 & 0 \\
$\Sigma^+$ & $\frac{1}{2}(N_c+1)$ & $\frac{1}{2}(N_c-3)$ & 1 &
$-\frac{1}{16} (N_c-3)(N_c+5)$ & $-\frac{1}{8}(N_c+5)$ &
$+\frac{1}{8}(N_c-3)$ \\
$\Sigma^0$ & $\frac{1}{2}(N_c-1)$ & $\frac{1}{2}(N_c-1)$ & 1 &
$-\frac{1}{16} (N_c^2 + 2N_c - 19)$ & $-\frac 1 2$ & $-\frac 1 2$ \\
$\Lambda$ & $\frac{1}{2}(N_c-1)$ & $\frac{1}{2}(N_c-1)$ & 1 &
$-\frac{1}{16} (N_c-1)(N_c+3)$ & 0 & 0 \\
$\Sigma^-$ & $\frac{1}{2}(N_c-3)$ & $\frac{1}{2}(N_c+1)$ & 1 &
$-\frac{1}{16} (N_c-3)(N_c+5)$ & $+\frac{1}{8} (N_c-3)$ &
$-\frac{1}{8} (N_c+5)$ \\
$\Xi^0$ & $\frac{1}{2}(N_c-1)$ & $\frac{1}{2}(N_c-3)$ & 2 &
$-\frac{1}{16} (N_c-3)(N_c+3)$ & $-\frac{1}{6} (N_c+3)$ &
$+\frac{1}{6} (N_c-3)$ \\
$\Xi^-$ & $\frac{1}{2}(N_c-3)$ & $\frac{1}{2}(N_c-1)$ & 2 &
$-\frac{1}{16} (N_c-3)(N_c+3)$ & $+\frac{1}{6} (N_c-3)$ &
$-\frac{1}{6} (N_c+3)$ \\
%
%
\hline
\end{tabular}
\end{table}

\begin{table}
\caption{Matrix elements of the operators $J_u^3$, $J_d^3$, and
$J_s^3$ in the state of maximal $J^3$.\label{S1}}
\begin{tabular}{c|c|c|c}
\hline\hline
State & $\langle J_u^3 \rangle$ & $\langle J_d^3 \rangle$ & $\langle
J_s^3 \rangle$ \\
\hline\hline
$\Delta^{++}$ & $+\frac{3}{20}(N_c+7)$ & $-\frac{3}{20}(N_c-3)$ & 0 \\
$\Delta^+$ & $+\frac{1}{20}(N_c+17)$ & $-\frac{1}{20}(N_c-13)$ & 0 \\
$\Delta^0$ & $-\frac{1}{20}(N_c-13)$ & $+\frac{1}{20}(N_c+17)$ & 0 \\
$\Delta^-$ & $-\frac{3}{20}(N_c-3)$ & $+\frac{3}{20}(N_c+7)$ & 0 \\
$\Sigma^{*+}$ & $+\frac{1}{8}(N_c+5)$ & $-\frac{1}{8}(N_c-3)$ &
$+\frac 1 2$ \\
$\Sigma^{*0}$ & $+\frac 1 2$ & $+\frac 1 2$ & $+\frac 1 2$ \\
$\Sigma^{*-}$ & $-\frac{1}{8}(N_c-3)$ & $+\frac{1}{8}(N_c+5)$ &
$+\frac 1 2$ \\
$\Xi^{*0}$ & $+\frac{1}{12}(N_c+3)$ & $-\frac{1}{12}(N_c-3)$ & $+1$ \\
$\Xi^{*-}$ & $-\frac{1}{12}(N_c-3)$ & $+\frac{1}{12}(N_c+3)$ & $+1$ \\
$\Omega^-$ & 0 & 0 & $+\frac 3 2$ \\
$p$ & $+\frac{1}{12}(N_c+5)$ & $-\frac{1}{12}(N_c-1)$ & 0 \\
$n$ & $-\frac{1}{12}(N_c-1)$ & $+\frac{1}{12}(N_c+5)$ & 0 \\
$\Sigma^+$ & $+\frac{1}{12}(N_c+5)$ & $-\frac{1}{12}(N_c-3)$ & $-\frac
1 6$ \\
$\Sigma^0$ & $+\frac 1 3$ & $+\frac 1 3$ & $-\frac 1 6$ \\
$\Lambda$ & 0 & 0 & $+\frac 1 2$ \\
$\Sigma^0 \Lambda$ & $-\frac{1}{12} \sqrt{(N_c-1)(N_c+3)}$ &
$+\frac{1}{12} \sqrt{(N_c-1)(N_c+3)}$ & 0 \\
$\Sigma^-$ & $-\frac{1}{12}(N_c-3)$ & $+\frac{1}{12}(N_c+5)$ & $-\frac
1 6$ \\
$\Xi^0$ & $-\frac{1}{36}(N_c+3)$ & $+\frac{1}{36}(N_c-3)$ & $+\frac 2
3$ \\
$\Xi^-$ & $+\frac{1}{36}(N_c-3)$ & $-\frac{1}{36}(N_c+3)$ & $+\frac 2
3$ \\
$\Delta^+ p$ & $+\frac{1}{6\sqrt{2}} \sqrt{(N_c-1)(N_c+5)}$ &
$-\frac{1}{6\sqrt{2}} \sqrt{(N_c-1)(N_c+5)}$ & 0 \\
$\Delta^0 n$ & $+\frac{1}{6\sqrt{2}} \sqrt{(N_c-1)(N_c+5)}$ &
$-\frac{1}{6\sqrt{2}} \sqrt{(N_c-1)(N_c+5)}$ & 0 \\
$\Sigma^{*0} \Lambda$ & $+\frac{1}{6\sqrt{2}} \sqrt{(N_c-1)(N_c+3)}$ &
$-\frac{1}{6\sqrt{2}} \sqrt{(N_c-1)(N_c+3)}$ & 0 \\
$\Sigma^{*0} \Sigma^0$ & $+\frac{1}{3\sqrt{2}}$ &
$+\frac{1}{3\sqrt{2}}$ & $-\frac{\sqrt{2}}{3}$ \\
$\Sigma^{*+} \Sigma^+$ & $+\frac{1}{12\sqrt{2}} (N_c+5)$ &
$-\frac{1}{12\sqrt{2}} (N_c-3)$ & $-\frac{\sqrt{2}}{3}$ \\
$\Sigma^{*-} \Sigma^-$ & $-\frac{1}{12\sqrt{2}} (N_c-3)$ &
$+\frac{1}{12\sqrt{2}} (N_c+5)$ & $-\frac{\sqrt{2}}{3}$ \\
$\Xi^{*0} \Xi^0$ & $+\frac{1}{9\sqrt{2}} (N_c+3)$ &
$-\frac{1}{9\sqrt{2}} (N_c-3)$ & $-\frac{\sqrt{2}}{3}$ \\
$\Xi^{*-} \Xi^-$ & $-\frac{1}{9\sqrt{2}} (N_c-3)$ &
$+\frac{1}{9\sqrt{2}} (N_c+3)$ & $-\frac{\sqrt{2}}{3}$ \\

\hline
\end{tabular}
\end{table}

\begin{table}
\caption{Matrix elements of the 1-, 2-, and 3-body operators 
corresponding to magnetic moments.\label{S2a}}
\begin{tabular}{c|c|c|c|c|c|c|c}
\hline\hline
State & $\langle J^3 \rangle$ & $\langle G^{33} \rangle$ & $\langle
G^{38} \rangle$ & $\langle T^3 J^3 \rangle$ & $\langle N_s J^3 \rangle$ 
& $\langle T^3 G^{33} \rangle$ & $\langle T^3 G^{38} \rangle$ \\
\hline\hline
$\Delta^{++}$ & $\frac{3}{2}$ & $\frac{3}{20}(N_c+2)$ 
& $\frac{\sqrt{3}}{4}$ & $\frac{9}{4}$ & 0 
& $\frac{9}{40}(N_c+2)$ & $\frac{3\sqrt{3}}{8}$ \\
$\Delta^+$ & $\frac{3}{2}$ & $\frac{1}{20}(N_c+2)$ 
& $\frac{\sqrt{3}}{4}$ & $\frac{3}{4}$ & 0 
& $\frac{1}{40}(N_c+2)$ & $\frac{\sqrt{3}}{8}$ \\
$\Delta^0$ & $\frac{3}{2}$ & $-\frac{1}{20}(N_c+2)$ 
& $\frac{\sqrt{3}}{4}$ & $-\frac{3}{4}$ & 0 
& $\frac{1}{40}(N_c+2)$ & $-\frac{\sqrt{3}}{8}$ \\
$\Delta^-$ & $\frac{3}{2}$ & $-\frac{3}{20}(N_c+2)$ 
& $\frac{\sqrt{3}}{4}$ & $-\frac{9}{4}$ & 0 
& $\frac{9}{40}(N_c+2)$ & $-\frac{3\sqrt{3}}{8}$ \\
$\Sigma^{*+}$ & $\frac{3}{2}$ & $\frac{1}{8}(N_c+1)$ 
& 0 & $\frac{3}{2}$ & $\frac{3}{2}$ 
& $\frac{1}{8}(N_c+1)$ & 0 \\
$\Sigma^{*0}$ & $\frac{3}{2}$ & 0 
& 0 & 0 & $\frac{3}{2}$ 
& 0 & 0  \\
$\Sigma^{*-}$ & $\frac{3}{2}$ & $-\frac{1}{8}(N_c+1)$ 
& 0 & $-\frac{3}{2}$ & $\frac{3}{2}$ 
& $\frac{1}{8}(N_c+1)$ & 0  \\
$\Xi^{*0}$ & $\frac{3}{2}$ & $\frac{1}{12}N_c$ 
& $-\frac{\sqrt{3}}{4}$ & $\frac{3}{4}$ 
& $3$ & $\frac{1}{24}N_c$ 
& $-\frac{\sqrt{3}}{8}$ \\
$\Xi^{*-}$ & $\frac{3}{2}$ & $-\frac{1}{12}N_c$ 
& $-\frac{\sqrt{3}}{4}$ & $-\frac{3}{4}$ 
& $3$ & $\frac{1}{24}N_c$ 
& $\frac{\sqrt{3}}{8}$ \\
$\Omega^-$ & $\frac{3}{2}$ & 0 & $-\frac{\sqrt{3}}{2}$ 
& 0 & $\frac{9}{2}$ & 0 & 0 \\
$p$ & $\frac{1}{2}$ & $\frac{1}{12}(N_c+2)$ & $\frac{1}{4\sqrt{3}}$ 
& $\frac{1}{4}$ & 0 & $\frac{1}{24}(N_c+2)$ 
& $\frac{1}{8\sqrt{3}}$ \\
$n$ & $\frac{1}{2}$ & $-\frac{1}{12}(N_c+2)$ & $\frac{1}{4\sqrt{3}}$ 
& $-\frac{1}{4}$ & 0 & $\frac{1}{24}(N_c+2)$ 
& $-\frac{1}{8\sqrt{3}}$ \\
$\Sigma^+$ & $\frac{1}{2}$ & $\frac{1}{12}(N_c+1)$ & $\frac{1}{2\sqrt{3}}$ 
& $\frac{1}{2}$ & $\frac{1}{2}$ & $\frac{1}{12}(N_c+1)$ 
& $\frac{1}{2\sqrt{3}}$ \\
$\Sigma^0$ & $\frac{1}{2}$ & 0 & $\frac{1}{2\sqrt{3}}$ 
& 0 & $\frac{1}{2}$ & 0 & 0 \\
$\Lambda$ & $\frac{1}{2}$ & 0 & $-\frac{1}{2\sqrt{3}}$ 
& 0 & $\frac{1}{2}$ & 0 & 0 \\
$\Sigma^0 \Lambda$ & 0 & $-\frac{1}{12}\sqrt{(N_c-1)(N_c+3)}$ 
& 0 & 0 & 0 & 0 & 0 \\
$\Sigma^-$ & $\frac{1}{2}$ & $-\frac{1}{12}(N_c+1)$ & $\frac{1}{2\sqrt{3}}$ 
& $-\frac{1}{2}$ & $\frac{1}{2}$ & $\frac{1}{12}(N_c+1)$ 
& $-\frac{1}{2\sqrt{3}}$ \\
$\Xi^0$ & $\frac{1}{2}$ & $-\frac{1}{36}N_c$ & $-\frac{\sqrt{3}}{4}$ 
& $\frac{1}{4}$ & $1$ & $-\frac{1}{72}N_c$ 
& $-\frac{\sqrt{3}}{8}$ \\
$\Xi^-$ & $\frac{1}{2}$ & $\frac{1}{36}N_c$ & $-\frac{\sqrt{3}}{4}$ 
& $-\frac{1}{4}$ & $1$ & $-\frac{1}{72}N_c$ 
& $\frac{\sqrt{3}}{8}$ \\
$\Delta^+ p$ & 0 & $\frac{1}{6\sqrt{2}}\sqrt{(N_c-1)(N_c+5)}$ 
& 0 & 0 & 0 & $\frac{1}{12\sqrt{2}}\sqrt{(N_c-1)(N_c+5)}$ 
& 0 \\
$\Delta^0 n$ & 0 & $\frac{1}{6\sqrt{2}}\sqrt{(N_c-1)(N_c+5)}$ 
& 0 & 0 & 0 & $-\frac{1}{12\sqrt{2}}\sqrt{(N_c-1)(N_c+5)}$ 
& 0 \\
$\Sigma^{*0} \Lambda$  & 0 & $\frac{1}{6\sqrt{2}}\sqrt{(N_c-1)(N_c+3)}$ 
& 0 & 0 & 0 & 0 & 0 \\
$\Sigma^{*0} \Sigma^0$  & 0 & 0 & $\frac{1}{\sqrt{6}}$ 
& 0 & 0 & 0 & 0 \\
$\Sigma^{*+} \Sigma^+$ & 0 & $\frac{1}{12\sqrt{2}}(N_c+1)$ 
& $\frac{1}{\sqrt{6}}$ & 0 & 0 & $\frac{1}{12\sqrt{2}}(N_c+1)$ 
& $\frac{1}{\sqrt{6}}$ \\
$\Sigma^{*-} \Sigma^-$ & 0 & $-\frac{1}{12\sqrt{2}}(N_c+1)$ 
& $\frac{1}{\sqrt{6}}$ & 0 & 0 & $\frac{1}{12\sqrt{2}}(N_c+1)$ 
& $-\frac{1}{\sqrt{6}}$ \\
$\Xi^{*0} \Xi^0$ & 0 & $\frac{1}{9\sqrt{2}}N_c$ 
& $\frac{1}{\sqrt{6}}$ & 0 & 0 & $\frac{1}{18\sqrt{2}}N_c$ 
& $\frac{1}{2\sqrt{6}}$ \\
$\Xi^{*-} \Xi^-$ & 0 & $-\frac{1}{9\sqrt{2}}N_c$ 
& $\frac{1}{\sqrt{6}}$ & 0 & 0 & $\frac{1}{18\sqrt{2}}N_c$ 
& $-\frac{1}{2\sqrt{6}}$ \\
\hline
\end{tabular}
\end{table}

\begin{table}
\caption{First continuation of above table.\label{S2b}}
\begin{tabular}{c|c|c|c|c|c}
\hline\hline
State & $\langle N_s G^{33} \rangle$ & $\langle N_s G^{38} \rangle$ & $\langle
J^2 J^3 \rangle$ & $\langle N_s^2 J^3 \rangle$ 
& $\langle (T^3)^2 J^3 \rangle$ \\
\hline\hline
$\Delta^{++}$ & 0 & 0 
& $\frac{45}{8}$ & 0 & $\frac{27}{8}$ \\
$\Delta^+$ & 0 & 0 
& $\frac{45}{8}$ & 0 & $\frac{3}{8}$  \\
$\Delta^0$ & 0 & 0 
& $\frac{45}{8}$ & 0 & $\frac{3}{8}$  \\
$\Delta^-$ & 0 & 0 
& $\frac{45}{8}$ & 0 & $\frac{27}{8}$  \\
$\Sigma^{*+}$ & $\frac{1}{8}(N_c+1)$  & 0 
& $\frac{45}{8}$ & $\frac{3}{2}$ & $\frac{3}{2}$ \\
$\Sigma^{*0}$ & 0 & 0 & $\frac{45}{8}$ 
& $\frac{3}{2}$ & 0 \\
$\Sigma^{*-}$ & $-\frac{1}{8}(N_c+1)$ & 0 
& $\frac{45}{8}$ & $\frac{3}{2}$ & $\frac{3}{2}$ \\
$\Xi^{*0}$ & $\frac{1}{6}N_c$ & $-\frac{\sqrt{3}}{2}$ 
& $\frac{45}{8}$ & $6$ & $\frac{3}{8}$ \\
$\Xi^{*-}$ & $-\frac{1}{6}N_c$ & $-\frac{\sqrt{3}}{2}$ 
& $\frac{45}{8}$ & $6$ & $\frac{3}{8}$ \\
$\Omega^-$ & 0 & $-\frac{3\sqrt{3}}{2}$ 
& $\frac{45}{8}$ & $\frac{27}{2}$ & 0 \\
$p$ & 0 & 0 
& $\frac{3}{8}$ & 0 & $\frac{1}{8}$ \\
$n$ & 0 & 0 
& $\frac{3}{8}$ & 0 & $\frac{1}{8}$  \\
$\Sigma^+$ & $\frac{1}{12}(N_c+1)$ & $\frac{1}{2\sqrt{3}}$ 
& $\frac{3}{8}$ & $\frac{1}{2}$ & $\frac{1}{2}$ \\
$\Sigma^0$ & 0 & $\frac{1}{2\sqrt{3}}$ 
& $\frac{3}{8}$ & $\frac{1}{2}$ & 0 \\
$\Lambda$ & 0 & $-\frac{1}{2\sqrt{3}}$ & $\frac{3}{8}$ 
& $\frac{1}{2}$ & 0 \\
$\Sigma^0 \Lambda$ & $-\frac{1}{12}\sqrt{(N_c-1)(N_c+3)}$ 
& 0 & 0 & 0 & 0 \\
$\Sigma^-$ & $-\frac{1}{12}(N_c+1)$ & $\frac{1}{2\sqrt{3}}$ 
& $\frac{3}{8}$ & $\frac{1}{2}$ & $\frac{1}{2}$ \\
$\Xi^0$ & $-\frac{1}{18}N_c$ & $-\frac{\sqrt{3}}{2}$ 
& $\frac{3}{8}$ & $2$ & $\frac{1}{8}$ \\
$\Xi^-$ & $\frac{1}{18}N_c$ & $-\frac{\sqrt{3}}{2}$ 
& $\frac{3}{8}$ & $2$ & $\frac{1}{8}$ \\
$\Delta^+ p$ & 0 & 0 
& 0 & 0 & 0 \\
$\Delta^0 n$ & 0 & 0 
& 0 & 0 & 0 \\
$\Sigma^{*0} \Lambda$ & $\frac{1}{6\sqrt{2}}\sqrt{(N_c-1)(N_c+3)}$ 
& 0 & 0 & 0 & 0 \\
$\Sigma^{*0} \Sigma^0$  & 0 & $\frac{1}{\sqrt{6}}$ 
& 0 & 0 & 0 \\
$\Sigma^{*+} \Sigma^+$ & $\frac{1}{12\sqrt{2}}(N_c+1)$ 
& $\frac{1}{\sqrt{6}}$ & 0 & 0 & 0 \\
$\Sigma^{*-} \Sigma^-$ & $-\frac{1}{12\sqrt{2}}(N_c+1)$ 
& $\frac{1}{\sqrt{6}}$ & 0 & 0 & 0 \\
$\Xi^{*0} \Xi^0$ & $\frac{\sqrt{2}}{9}N_c$ 
& $\sqrt{\frac{2}{3}}$ & 0 & 0 & 0 \\
$\Xi^{*-} \Xi^-$ & $-\frac{\sqrt{2}}{9}N_c$ 
& $\sqrt{\frac{2}{3}}$ & 0 & 0 & 0 \\
\hline
\end{tabular}
\end{table}
\begin{table}
\caption{Second continuation of above table.\label{S2c}}
\begin{tabular}{c|c|c|c|c}
\hline\hline
State & $\langle T^3 N_s J^3 \rangle$ 
& $\frac{1}{2} \langle \{ J^2 , G^{33} \} \rangle$ 
& $\frac{1}{2} \langle \{ J^2 , G^{38} \} \rangle$  
& $\langle (T^3)^2 G^{33} \rangle$ \\
\hline\hline
$\Delta^{++}$ & 0 & $\frac{9}{16}(N_c+2)$ 
& $\frac{15\sqrt{3}}{16}$ & $\frac{27}{80}(N_c+2)$ \\
$\Delta^+$ & 0 & $\frac{3}{16}(N_c+2)$ 
& $\frac{15\sqrt{3}}{16}$ & $\frac{1}{80}(N_c+2)$ \\
$\Delta^0$ & 0 & $-\frac{3}{16}(N_c+2)$ 
& $\frac{15\sqrt{3}}{16}$ & $-\frac{1}{80}(N_c+2)$ \\
$\Delta^-$ & 0 & $-\frac{9}{16}(N_c+2)$ 
& $\frac{15\sqrt{3}}{16}$ & $-\frac{27}{80}(N_c+2)$ \\
$\Sigma^{*+}$ & $\frac{3}{2}$ & $\frac{15}{32}(N_c+1)$ 
& 0 & $\frac{1}{8}(N_c+1)$ \\
$\Sigma^{*0}$ & 0 & 0 & 0 & 0 \\
$\Sigma^{*-}$ & $-\frac{3}{2}$ & $-\frac{15}{32}(N_c+1)$ 
& 0 & $-\frac{1}{8}(N_c+1)$ \\
$\Xi^{*0}$ & $\frac{3}{2}$ & $\frac{5}{16}N_c$ 
& $-\frac{15\sqrt{3}}{16}$ & $\frac{1}{48}N_c$ \\
$\Xi^{*-}$ & $-\frac{3}{2}$ & $-\frac{5}{16}N_c$ 
& $-\frac{15\sqrt{3}}{16}$ & $-\frac{1}{48}N_c$ \\
$\Omega^-$ & 0 & 0 & $-\frac{15\sqrt{3}}{8}$ & 0 \\
$p$ & 0 & $\frac{1}{16}(N_c+2)$ 
& $\frac{\sqrt{3}}{16}$ & $\frac{1}{48}(N_c+2)$ \\
$n$ & 0 & $-\frac{1}{16}(N_c+2)$ 
& $\frac{\sqrt{3}}{16}$ & $-\frac{1}{48}(N_c+2)$ \\
$\Sigma^+$ & $\frac{1}{2}$ & $\frac{1}{16}(N_c+1)$ 
& $\frac{\sqrt{3}}{8}$ & $\frac{1}{12}(N_c+1)$ \\
$\Sigma^0$ & 0 & 0 & $\frac{\sqrt{3}}{8}$ & 0 \\
$\Lambda$ & 0 & 0 & $-\frac{\sqrt{3}}{8}$ & 0 \\
$\Sigma^0 \Lambda$ & 0 & $-\frac{1}{16}\sqrt{(N_c-1)(N_c+3)}$ & 0 & 0 \\
$\Sigma^-$ & $-\frac{1}{2}$ & $-\frac{1}{16}(N_c+1)$ 
& $\frac{\sqrt{3}}{8}$ & $-\frac{1}{12}(N_c+1)$ \\
$\Xi^0$ & $\frac{1}{2}$ & $-\frac{1}{48}N_c$ 
& $-\frac{3\sqrt{3}}{16}$ & $-\frac{1}{144}N_c$ \\
$\Xi^-$  & $-\frac{1}{2}$ & $\frac{1}{48}N_c$ 
& $-\frac{3\sqrt{3}}{16}$ & $\frac{1}{144}N_c$ \\
$\Delta^+ p$ & 0 & $\frac{3}{8\sqrt{2}}\sqrt{(N_c-1)(N_c+5)}$ & 0 
& $\frac{1}{24\sqrt{2}}\sqrt{(N_c-1)(N_c+5)}$ \\
$\Delta^0 n$ & 0 & $\frac{3}{8\sqrt{2}}\sqrt{(N_c-1)(N_c+5)}$ & 0 
& $\frac{1}{24\sqrt{2}}\sqrt{(N_c-1)(N_c+5)}$ \\
$\Sigma^{*0} \Lambda$  & 0 & $\frac{3}{8\sqrt{2}}\sqrt{(N_c-1)(N_c+3)}$ 
& 0 & 0 \\
$\Sigma^{*0} \Sigma^0$ & 0 & 0 & $\frac{3}{4}\sqrt{\frac{3}{2}}$ & 0 \\
$\Sigma^{*+} \Sigma^+$ & 0 & $\frac{3}{16\sqrt{2}}(N_c+1)$ 
& $\frac{3}{4}\sqrt{\frac{3}{2}}$ & $\frac{1}{12\sqrt{2}}(N_c+1)$ \\
$\Sigma^{*-} \Sigma^-$ & 0 & $-\frac{3}{16\sqrt{2}}(N_c+1)$ 
& $\frac{3}{4}\sqrt{\frac{3}{2}}$ & $-\frac{1}{12\sqrt{2}}(N_c+1)$ \\
$\Xi^{*0} \Xi^0$ & 0 & $\frac{1}{4\sqrt{2}}N_c$ 
& $\frac{3}{4}\sqrt{\frac{3}{2}}$ & $\frac{1}{36\sqrt{2}}N_c$ \\
$\Xi^{*-} \Xi^-$ & 0 & $-\frac{1}{4\sqrt{2}}N_c$ 
& $\frac{3}{4}\sqrt{\frac{3}{2}}$ & $-\frac{1}{36\sqrt{2}}N_c$ \\
\hline
\end{tabular}
\end{table}
\begin{table}
\caption{Third continuation of above table.\label{S2d}}
\begin{tabular}{c|c|c|c|c}
\hline\hline
State & $\langle (T^3)^2 G^{38} \rangle$ 
& $\langle N_s^2 G^{33} \rangle$ 
& $\langle N_s^2 G^{38}  \rangle$  
& $\langle T^3 N_s G^{33} \rangle$ \\
\hline\hline
$\Delta^{++}$ & $\frac{9\sqrt{3}}{16}$ & 0 
& 0 & 0 \\
$\Delta^+$ & $\frac{\sqrt{3}}{16}$ & 0 
& 0 & 0 \\
$\Delta^0$ & $\frac{\sqrt{3}}{16}$ & 0 
& 0 & 0 \\
$\Delta^-$ & $\frac{9\sqrt{3}}{16}$ & 0 
& 0 & 0 \\
$\Sigma^{*+}$ & 0 & $\frac{1}{8}(N_c+1)$ & 0 
& $\frac{1}{8}(N_c+1)$ \\
$\Sigma^{*0}$ & 0 & 0 & 0 & 0 \\
$\Sigma^{*-}$ & 0 & $-\frac{1}{8}(N_c+1)$ & 0 
& $\frac{1}{8}(N_c+1)$ \\
$\Xi^{*0}$ & $-\frac{\sqrt{3}}{16}$ & $\frac{1}{3}N_c$ 
& $-\sqrt{3}$ & $\frac{1}{12}N_c$ \\
$\Xi^{*-}$ & $-\frac{\sqrt{3}}{16}$ & $-\frac{1}{3}N_c$ 
& $-\sqrt{3}$ & $\frac{1}{12}N_c$ \\
$\Omega^-$ & 0 & 0 & $-\frac{9\sqrt{3}}{2}$ & 0 \\
$p$ & $\frac{1}{16\sqrt{3}}$ & 0 
& 0 & 0 \\
$n$ & $\frac{1}{16\sqrt{3}}$ & 0 
& 0 & 0 \\
$\Sigma^+$ & $\frac{1}{2\sqrt{3}}$ & $\frac{1}{12}(N_c+1)$ 
& $\frac{1}{2\sqrt{3}}$ & $\frac{1}{12}(N_c+1)$ \\
$\Sigma^0$ & 0 & 0 & $\frac{1}{2\sqrt{3}}$ & 0 \\
$\Lambda$ & 0 & 0 & $-\frac{1}{2\sqrt{3}}$ & 0 \\
$\Sigma^0 \Lambda$ & 0 & $-\frac{1}{12}\sqrt{(N_c-1)(N_c+3)}$ 
& 0 & 0 \\
$\Sigma^-$ & $\frac{1}{2\sqrt{3}}$ & $-\frac{1}{12}(N_c+1)$ 
& $\frac{1}{2\sqrt{3}}$ & $\frac{1}{12}(N_c+1)$ \\
$\Xi^0$ & $-\frac{\sqrt{3}}{16}$ & $-\frac{1}{9}N_c$ 
& $-\sqrt{3}$ & $-\frac{1}{36}N_c$ \\
$\Xi^-$ & $-\frac{\sqrt{3}}{16}$ & $\frac{1}{9}N_c$ 
& $-\sqrt{3}$ & $-\frac{1}{36}N_c$ \\
$\Delta^+ p$ & 0 & 0 
& 0 & 0 \\
$\Delta^0 n$ & 0 & 0 
& 0 & 0 \\
$\Sigma^{*0} \Lambda$  & 0 
& $\frac{1}{6\sqrt{2}}\sqrt{(N_c-1)(N_c+3)}$ 
& 0 & 0 \\
$\Sigma^{*0} \Sigma^0$ & 0 & 0 & $\frac{1}{\sqrt{6}}$ & 0 \\
$\Sigma^{*+} \Sigma^+$ & $\frac{1}{\sqrt{6}}$ 
& $\frac{1}{12\sqrt{2}}(N_c+1)$ & $\frac{1}{\sqrt{6}}$ 
& $\frac{1}{12\sqrt{2}}(N_c+1)$ \\
$\Sigma^{*-} \Sigma^-$ & $\frac{1}{\sqrt{6}}$ 
& $-\frac{1}{12\sqrt{2}}(N_c+1)$ & $\frac{1}{\sqrt{6}}$ 
& $\frac{1}{12\sqrt{2}}(N_c+1)$ \\
$\Xi^{*0} \Xi^0$ & $\frac{1}{4\sqrt{6}}$ 
& $\frac{2\sqrt{2}}{9}N_c$ & $\frac{2\sqrt{2}}{\sqrt{3}}$ 
& $\frac{1}{9\sqrt{2}}N_c$ \\
$\Xi^{*-} \Xi^-$ & $\frac{1}{4\sqrt{6}}$ 
& $-\frac{2\sqrt{2}}{9}N_c$ & $\frac{2\sqrt{2}}{\sqrt{3}}$ 
& $\frac{1}{9\sqrt{2}}N_c$ \\
\hline
\end{tabular}
\end{table}
\begin{table}
\caption{Fourth continuation of above table.\label{S2e}}
\begin{tabular}{c|c|c|c|c}
\hline\hline
State & $\langle T^3 N_s G^{38} \rangle$ 
& $\langle (J^i G^{i3}) J^3 \rangle$ 
& $\langle (J^i G^{i8}) J^3  \rangle$  
& $\frac{1}{2} \langle \{ J^i G^{i8},G^{33} \}  \rangle$ \\
\hline\hline
$\Delta^{++}$ & 0 & $\frac{9}{16}(N_c+2)$ 
& $\frac{15\sqrt{3}}{16}$ & $\frac{3\sqrt{3}}{32}(N_c+2)$ \\
$\Delta^+$ & 0 & $\frac{3}{16}(N_c+2)$ 
& $\frac{15\sqrt{3}}{16}$ & $\frac{\sqrt{3}}{32}(N_c+2)$ \\
$\Delta^0$ & 0 & $-\frac{3}{16}(N_c+2)$ 
& $\frac{15\sqrt{3}}{16}$ & $-\frac{\sqrt{3}}{32}(N_c+2)$ \\
$\Delta^-$ & 0 & $-\frac{9}{16}(N_c+2)$ 
& $\frac{15\sqrt{3}}{16}$ & $-\frac{3\sqrt{3}}{32}(N_c+2)$ \\
$\Sigma^{*+}$ & 0 & $\frac{15}{32}(N_c+1)$ & 0 & 0 \\
$\Sigma^{*0}$ & 0 & 0 & 0 & 0 \\
$\Sigma^{*-}$ & 0 & $-\frac{15}{32}(N_c+1)$ & 0 & 0 \\
$\Xi^{*0}$ & $-\frac{\sqrt{3}}{4}$ & $\frac{5}{16}N_c$ 
& $-\frac{15\sqrt{3}}{16}$ & $-\frac{5}{32\sqrt{3}}N_c$ \\
$\Xi^{*-}$ & $\frac{\sqrt{3}}{4}$ & $-\frac{5}{16}N_c$ 
& $-\frac{15\sqrt{3}}{16}$ & $\frac{5}{32\sqrt{3}}N_c$ \\
$\Omega^-$ & 0 & 0 & $-\frac{15\sqrt{3}}{8}$ & 0 \\
$p$ & 0 & $\frac{1}{16}(N_c+2)$ 
& $\frac{\sqrt{3}}{16}$ & $\frac{1}{32\sqrt{3}}(N_c+2)$ \\
$n$ & 0 & $-\frac{1}{16}(N_c+2)$ 
& $\frac{\sqrt{3}}{16}$ & $-\frac{1}{32\sqrt{3}}(N_c+2)$ \\
$\Sigma^+$ & $\frac{1}{2\sqrt{3}}$ & $\frac{1}{16}(N_c+1)$ 
& $\frac{\sqrt{3}}{8}$ & $\frac{1}{16\sqrt{3}}(N_c+1)$ \\
$\Sigma^0$ & 0 & 0 & $\frac{\sqrt{3}}{8}$ & 0 \\
$\Lambda$ & 0 & 0 & $-\frac{\sqrt{3}}{8}$ & 0 \\
$\Sigma^0 \Lambda$ & 0 & $-\frac{1}{16}\sqrt{(N_c-1)(N_c+3)}$ & 0 & 0 \\
$\Sigma^-$ & $-\frac{1}{2\sqrt{3}}$ & $-\frac{1}{16}(N_c+1)$ 
& $\frac{\sqrt{3}}{8}$ & $-\frac{1}{16\sqrt{3}}(N_c+1)$ \\
$\Xi^0$ & $-\frac{\sqrt{3}}{4}$ & $-\frac{1}{48}N_c$ 
& $-\frac{3\sqrt{3}}{16}$ & $\frac{1}{32\sqrt{3}}N_c$ \\
$\Xi^-$ & $\frac{\sqrt{3}}{4}$ & $\frac{1}{48}N_c$ 
& $-\frac{3\sqrt{3}}{16}$ & $-\frac{1}{32\sqrt{3}}N_c$ \\
$\Delta^+ p$ & 0 & 0 & 0 
& $\frac{1}{16}\sqrt{\frac{3}{2}}\sqrt{(N_c-1)(N_c+5)}$ \\
$\Delta^0 n$ & 0 & 0 & 0 
& $\frac{1}{16}\sqrt{\frac{3}{2}}\sqrt{(N_c-1)(N_c+5)}$ \\
$\Sigma^{*0} \Lambda$  & 0 & 0 & 0 
& $-\frac{1}{16\sqrt{6}}\sqrt{(N_c-1)(N_c+3)}$ \\
$\Sigma^{*0} \Sigma^0$ & 0 & 0 & 0 & 0 \\
$\Sigma^{*+} \Sigma^+$ & $\frac{1}{\sqrt{6}}$ & 0 & 0 
& $\frac{1}{32\sqrt{6}}(N_c+1)$ \\
$\Sigma^{*-} \Sigma^-$ & $-\frac{1}{\sqrt{6}}$ & 0 & 0 
& $-\frac{1}{32\sqrt{6}}(N_c+1)$ \\
$\Xi^{*0} \Xi^0$ & $\frac{1}{\sqrt{6}}$ & 0 & 0 
& $-\frac{1}{6\sqrt{6}}N_c$ \\
$\Xi^{*-} \Xi^-$ & $-\frac{1}{\sqrt{6}}$ & 0 & 0 
& $\frac{1}{6\sqrt{6}}N_c$ \\
\hline
\end{tabular}
\end{table}
\begin{table}
\caption{Fifth continuation of above table.\label{S2f}}
\begin{tabular}{c|c|c|c}
\hline\hline
State & $\frac{1}{2} \langle \{ J^i G^{i8},G^{38} \} \rangle$ 
& $\frac{1}{2} \langle \{ J^i G^{i3},G^{33} \} \rangle$ 
& $\frac{1}{2} \langle \{ J^i G^{i3},G^{38} \} \rangle$ \\
\hline\hline
$\Delta^{++}$ & $\frac{15}{32}$ & $\frac{9}{160}(N_c+2)^2$ 
& $\frac{3\sqrt{3}}{32}(N_c+2)$ \\
$\Delta^+$ & $\frac{15}{32}$ & $\frac{1}{160}(N_c+2)^2$ 
& $\frac{\sqrt{3}}{32}(N_c+2)$ \\
$\Delta^0$ & $\frac{15}{32}$ & $\frac{1}{160}(N_c+2)^2$ 
& $-\frac{\sqrt{3}}{32}(N_c+2)$ \\
$\Delta^-$ & $\frac{15}{32}$ & $\frac{9}{160}(N_c+2)^2$ 
& $-\frac{3\sqrt{3}}{32}(N_c+2)$ \\
$\Sigma^{*+}$ & 0 & $\frac{5}{128}(N_c+1)^2$  & 0 \\
$\Sigma^{*0}$ & 0 & 0 & 0 \\
$\Sigma^{*-}$ & 0 & $\frac{5}{128}(N_c+1)^2$ & 0 \\
$\Xi^{*0}$ & $\frac{15}{32}$ & $\frac{5}{288}N_c^2$ 
& $-\frac{5}{32\sqrt{3}}N_c$  \\
$\Xi^{*-}$ & $\frac{15}{32}$ & $\frac{5}{288}N_c^2$ 
& $\frac{5}{32\sqrt{3}}N_c$  \\
$\Omega^-$ & $\frac{15}{8}$ & 0 & 0 \\
$p$ & $\frac{1}{32}$ & $\frac{1}{96}(N_c+2)^2$ 
& $\frac{1}{32\sqrt{3}}(N_c+2)$ \\
$n$ & $\frac{1}{32}$ & $\frac{1}{96}(N_c+2)^2$ 
& $-\frac{1}{32\sqrt{3}}(N_c+2)$ \\
$\Sigma^+$ & $\frac{1}{8}$ & $\frac{1}{96}(N_c+1)^2$ 
& $\frac{1}{16\sqrt{3}}(N_c+1)$ \\
$\Sigma^0$ & $\frac{1}{8}$ & $\frac{1}{96}(N_c-1)(N_c+3)$ & 0 \\
$\Lambda$ & $\frac{1}{8}$ & $\frac{1}{96}(N_c-1)(N_c+3)$ & 0 \\
$\Sigma^0 \Lambda$ & 0 & 0 & 0 \\
$\Sigma^-$ & $\frac{1}{8}$ & $\frac{1}{96}(N_c+1)^2$ 
& $-\frac{1}{16\sqrt{3}}(N_c+1)$ \\
$\Xi^0$ & $\frac{9}{32}$ & $\frac{1}{864}N_c^2$ 
& $\frac{1}{32\sqrt{3}}N_c$ \\
$\Xi^-$ & $\frac{9}{32}$ & $\frac{1}{864}N_c^2$ 
& $-\frac{1}{32\sqrt{3}}N_c$ \\
$\Delta^+ p$ & 0 & $\frac{1}{48\sqrt{2}}(N_c+2)\sqrt{(N_c-1)(N_c+5)}$ & 0 \\
$\Delta^0 n$ & 0 & $-\frac{1}{48\sqrt{2}}(N_c+2)\sqrt{(N_c-1)(N_c+5)}$ & 0 \\
$\Sigma^{*0} \Lambda$  & 0 & 0 &
$-\frac{1}{16\sqrt{6}} \sqrt{(N_c-1)(N_c+3)}$ \\
$\Sigma^{*0} \Sigma^0$ & $\frac{1}{8\sqrt{2}}$ & 0 & 0 \\
$\Sigma^{*+} \Sigma^+$ & $\frac{1}{8\sqrt{2}}$ 
& $\frac{7}{384\sqrt{2}}(N_c+1)^2$ & $\frac{7}{32\sqrt{6}}(N_c+1)$ \\
$\Sigma^{*-} \Sigma^-$ & $\frac{1}{8\sqrt{2}}$ 
& $\frac{7}{384\sqrt{2}}(N_c+1)^2$ & $-\frac{7}{32\sqrt{6}}(N_c+1)$ \\
$\Xi^{*0} \Xi^0$ & $-\frac{1}{2\sqrt{2}}$ & $\frac{1}{108\sqrt{2}}N_c^2$ 
& $\frac{1}{12\sqrt{6}}N_c$ \\
$\Xi^{*-} \Xi^-$ & $-\frac{1}{2\sqrt{2}}$ & $\frac{1}{108\sqrt{2}}N_c^2$ 
& $-\frac{1}{12\sqrt{6}}N_c$ \\
\hline
\end{tabular}
\end{table}
\begin{table}
\caption{Best fit values for the coefficients in the expansion
Eq.~(\ref{newexp}).\label{fit}}
\begin{tabular}{llll}
\hline\hline
$d_1 = +0.987 \pm 0.038$ \ & $d_2 = +0.076 \pm 0.092$ \ &
$d_3 = +1.385 \pm 0.258$ \ & $d_4 = +0.059 \pm 0.255$ \\
$d_5 = -0.348 \pm 0.114$ \ & $d_6 = -0.140 \pm 0.108$ \ &
$d_7 = +0.126 \pm 0.108$ \\
\hline
\end{tabular}
\end{table}
\begin{table}
\caption{Best fit values for the 16 unknown magnetic moments in units
of $\mu_N$.\label{pred}}
\begin{tabular}{rrrr}
\hline\hline
$\mu_{\Delta^{\! +}}    = +3.04 \pm 0.13$ & \
$\mu_{\Delta^{\! 0}}    = +0.00 \pm 0.10$ & \
$\mu_{\Delta^{\! -}}    = -3.04 \pm 0.13$ & \
$\mu_{\Sigma^{*+}}      = +3.35 \pm 0.13$ \\
$\mu_{\Sigma^{*0}}      = +0.32 \pm 0.11$ & \
$\mu_{\Sigma^{*-}}      = -2.70 \pm 0.13$ & \
$\mu_{\Xi^{*0}}         = +0.64 \pm 0.11$ & \
$\mu_{\Xi^{*-}}         = -2.36 \pm 0.14$ \\
$\mu_{\Sigma^0}         = +0.77 \pm 0.10$ & \
$\mu_{\Delta^{\! 0} n}  = +3.51 \pm 0.11$ & \
$\mu_{\Sigma^{*0} \Lambda}  = +2.93 \pm 0.11$ & \
$\mu_{\Sigma^{*0} \Sigma^0} = +1.39 \pm 0.11$ \\
$\mu_{\Sigma^{*+} \Sigma^+} = +2.97 \pm 0.11$ & \
$\mu_{\Sigma^{*-} \Sigma^-} = -0.19 \pm 0.11$ & \
$\mu_{\Xi^{*0} \Xi^0}       = +2.96 \pm 0.12$ & \
$\mu_{\Xi^{*-} \Xi^-}       = -0.19 \pm 0.11$ \\
\hline
\end{tabular}
\end{table}

\begin{thebibliography}{99}

\bibitem{reviews}
R.F.~Lebed, Czech.\ J. Nucl.\ Phys.\ {\bf 49}, 1273 (1999),
nucl-th/9810080;
E.~Jenkins, Ann.\ Rev.\ Nucl.\ Part.\ Sci.\ {\bf 48}, 81 (1998),
hep-ph/9803349;
A.V.~Manohar, in {\it Probing the Standard Model of Particle
Interactions}, ed.\ by R.~Gupta {\it et al.}  (Elsevier, Amsterdam,
1999), hep-ph/9802419.
Note that the newest of these is already five years old, and many more
recent advances have occurred since then.

\bibitem{DM}
R.F.~Dashen and A.V.~Manohar, Phys.\ Lett.\ B {\bf 315}, 425 (1993).

\bibitem{DJM1}
R.F.~Dashen, E.~Jenkins, and A.V.~Manohar, Phys.\ Rev.\ D {\bf 49},
4713 (1994); D {\bf 51}, 2489(E) (1995).

\bibitem{Jenk}
E.~Jenkins, Phys.\ Lett.\ B {\bf 315}, 441 (1993).

\bibitem{JM}
E.~Jenkins and A.V.~Manohar, Phys.\ Lett.\ B {\bf 335}, 452 (1994).

\bibitem{LMRW}
M.A.~Luty, J.~March-Russell, and M.~White, Phys.\ Rev.\ D {\bf 51},
2332 (1994).

\bibitem{DDJM}
J.~Dai, R.F.~Dashen, E.~Jenkins, and A.V.~Manohar, Phys.\ Rev.\ D {\bf
53}, 273 (1996).

\bibitem{RLSU6}
R.F.~Lebed, Phys.\ Rev.\ D {\bf 51}, 5039 (1995).

\bibitem{BH}
A.J.~Buchmann and E.M.~Henley, Phys.\ Rev.\ D {\bf 65}, 073017 (2002).

\bibitem{JL}
E.~Jenkins and R.F.~Lebed, Phys.\ Rev.\ D {\bf 52}, 282 (1995);
E.~Jenkins and R.F.~Lebed, Phys.\ Rev.\ D {\bf 62}, 077901 (2000);
P.F.~Bedaque and M.A.~Luty, Phys.\ Rev.\ D {\bf 54}, 2317 (1996).

\bibitem{BL1}
A.J.~Buchmann and R.F.~Lebed, Phys.\ Rev.\ D {\bf 62}, 096005 (2000).

\bibitem{BHL}
A.J.~Buchmann, J.A.~Hester, and R.F.~Lebed, Phys.\ Rev.\ D {\bf 66},
056002 (2002).

\bibitem{BL2}
A.J.~Buchmann and R.F.~Lebed, Phys.\ Rev.\ D {\bf 67}, 016002 (2003).

\bibitem{Goity}
J.L.~Goity, Phys.\ Lett.\ B {\bf 414}, 140 (1997).

\bibitem{CCGL}
C.E.~Carlson, C.D.~Carone, J.L.~Goity, and R.F.~Lebed, Phys.\ Lett.\ B
{\bf 438}, 327 (1998);
Phys.\ Rev.\ D {\bf 59}, 114008 (1999).

\bibitem{CC}
C.E.~Carlson and C.D.~Carone, Phys.\ Rev.\ D {\bf 58}, 053005 (1998);
Phys.\ Lett.\ B {\bf 441}, 363 (1998);
Phys.\ Lett.\ B {\bf 260}, (2000).

\bibitem{GSS}
J.L.~Goity, C.L.~Schat, and N.N.~Scoccola, Phys.\ Rev.\ Lett.\ {\bf
88}, 102002 (2002);
Phys.\ Rev.\ D {\bf 66}, 114014 (2002);
Phys.\ Lett.\ B {\bf 564}, 83 (2003).

\bibitem{PS}
D.~Pirjol and C.~Schat, Phys.\ Rev.\ D {\bf 67}, 096009 (2003).

\bibitem{JJM}
E.~Jenkins, X.~Ji, and A.V.~Manohar, Phys.\ Rev.\ Lett.\ {\bf 89},
242001 (2002).

\bibitem{DJM2}
R.F.~Dashen, E.~Jenkins, and A.V.~Manohar, Phys.\ Rev.\ D {\bf 51},
3697 (1995).

\bibitem{edmonds}
A.R.~Edmonds, {\it Angular Momentum in Quantum Mechanics} (Princeton
Univ.\ Press, Princeton, NJ, 1996).

\bibitem{PDG}
Particle Data Group (K.~Hagiwara {\it et al.}), Phys.\ Rev.\ D {\bf
66}, 010001 (2002).

\bibitem{LCM}
G.~L\'{o}pez~Castro and A.~Mariano, Phys.\ Lett.\ B {\bf 517}, 339
(2001); Nucl.\ Phys.\ {\bf A697}, 440 (2002).

\bibitem{Kot}
M.~Kotulla {\it et al.}, Phys.\ Rev.\ Lett.\ {\bf 89}, 272001 (2002).

\bibitem{CohenWZ}
T.D.~Cohen, hep-ph/0312191 and references therein.

\bibitem{JLSM}
E.~Jenkins, M.E.~Luke, M.J.~Savage, and A.V.~Manohar, Phys.\ Lett.\ B
{\bf 302}, 482 (1993); B {\bf 388}, 866(E) (1996).

\bibitem{Meis72}
G.W.~Meisner {\it et al.}, Nuovo Cimento A {\bf 12}, 62 (1972).

\end{thebibliography}
\end{document}